\documentclass[12pt]{article}

\usepackage{amssymb,amsmath,amsthm}
\usepackage{hyperref}
\usepackage{array,longtable}
\usepackage[dvips]{graphicx}
\usepackage[dvips]{color}

\newcommand{\version}{December 14, 2004}

\usepackage[mathscr]{eucal}

\newcommand{\finkfile}{}

\newcommand{\Ds}{\mathscr{D}}

\newcommand{\Rh}{\mathbb R}
\newcommand{\Ch}{\mathbb C}
\newcommand{\Zh}{\mathbb Z}

\newcommand{\Nc}{\mathcal{N}}

\newcommand{\Mc}{\mathcal{M}}

\newcommand{\pd}{\partial}

\newcommand{\Tr}{\mathop{\mathrm{Tr}}\nolimits}

\newcommand{\vol}{\mathop{\mathrm{vol}}\nolimits}

\renewcommand{\Im}{\mathop{\mathrm{Im}}\nolimits}

\renewcommand{\theequation}{\arabic{section}.\arabic{equation}}

\begin{document}

\title{
\begin{flushright}
{\small hep-th/0412167}
\end{flushright}
\vspace{1cm}
Conformal blocks for $AdS_5$ singletons}
\author{
{Dmitriy ~Belov\footnote{Email: \texttt{belov@physics.rutgers.edu}}
\; and
\;Gregory W.~Moore\footnote{Email: \texttt{gmoore@physics.rutgers.edu}}}
\vspace{5mm}
\\
\emph{
Department of Physics, Rutgers University}
\\
\emph{136 Frelinghuysen Rd., Piscataway, NJ 08854, USA}
}

\date{\version}

\maketitle

\thispagestyle{empty}

\begin{abstract}
We give a simple  derivation of the conformal blocks of the
singleton sector of compactifications of IIB string theory
on spacetimes of the form $X_5\times Y_5$ with
$Y_5$ compact, while $X_5$ has as conformal boundary an arbitrary
$4$-manifold $M_4$.  We retain the second-derivative terms in
the action for the $B,C$ fields and thus the analysis is not
purely topological. The unit-normalized conformal blocks agree
exactly with the quantum partition function of the $U(1)$ gauge
theory on the conformal boundary. We reproduce the action of the
magnetic translation group and the $SL(2,\Zh)$ $S$-duality group obtained
from the purely topological analysis of Witten.
An interesting subtlety in the normalization of the IIB
Chern-Simons phase is noted.

\end{abstract}

\clearpage

\tableofcontents
\clearpage

\section{Introduction and Conclusion }
\label{sec:intro}
\dopage{\finkfile}
\setcounter{equation}{0}

In AdS compactifications of string theory and M-theory there is a free field sector of
the theory known as the singleton sector. In the bulk description these
 are   typically gauge modes,  which do not propagate in the interior but do become
dynamical on the conformal boundary, thanks to the  Chern-Simons terms in the
supergravity action. Despite the fact that the singleton sector is
``just a free field theory'' it has been a source of some confusion. In this
paper we give a simple and straightforward derivation of the conformal blocks of the
singleton sector for compactifications of type IIB strings on spacetimes of the
form $X_5 \times Y_5$, where $Y_5$ is compact, while $X_5$ is noncompact with
a conformal boundary $M_4$. The blocks depend on  the topology of
$M_4$ in an interesting way, so we assume $M_4$ is a general compact $4$-manifold.
There is an $S$-duality anomaly if $M_4$ is not spin. This was observed in
\cite{Witten:1998wy} and we reproduce the result in section 3.

The singleton sector was first studied by Witten in \cite{Witten:1998wy} based on the
topological field theory with exponentiated action
\begin{equation}
\exp\bigl[ 2\pi i N \int_{X_5} B_2 d C_2\bigr].
\label{toplag}
\end{equation}
Here $N$ is the 5-form flux through $X_5$ and $B_2,C_2$ are the
supergravity potentials with fieldstrengths $H_3, F_3$. Our results
are in accord with \cite{Witten:1998wy}, but in the present paper we retain the
second derivative terms in the action. This leads to some differences in the
analysis of the conformal blocks. Moreover,   as stressed in \cite{Maldacena:2001ss,Gukov:2004id}
the Hamiltonian governing the dynamics of  the
singleton modes is not determined
by the Chern-Simons action alone. The  method we use determines the
Hamiltonian for the singleton modes, and   allows us to solve explicitly
for the conformal blocks of the singleton sector in terms of $\Theta$-functions.
The singleton sector is holographically dual to a free ${\cal N}=4$ supersymmetric
theory with gauge group $U(1)$.
Our main result is summarized by the Lagrangian for the $U(1)$ gauge boson.
The Lagrangian summarizes the coupling of the gauge boson to the harmonic
modes of $B_2, C_2$ at the conformal boundary.
It depends on the topological sector $\beta \in H^2(M_4, \Zh/N\Zh) $
and is given by equation $(4.23)$ below. The bulk supergravity interpretation
of $\beta$ is that it is a ``Page charge'' for the $(B_2,C_2)$ system,
much as in \cite{mooreS04}. It is quite curious that requiring that the conformal
blocks be properly normalized in their natural inner product correctly
reproduces the one-loop determinants of the $U(1)$ gauge theory  on the boundary. We show this in
section 4.3.

The full partition function of the string theory on $X_5 \times Y_5$ will be of the form
\begin{equation}
\sum_\beta Z^\beta Z^{\rm singleton}_\beta
\label{decomp}
\end{equation}
Here $Z^{\rm singleton}_\beta$ are the conformal blocks derived in this paper.
They are functions of $\tau$ and of the harmonic modes of $B_2,C_2$
on the conformal boundary. The
dependence of the partition function on the remaining boundary values of IIB supergravity
fields enter into $Z^\beta$. These will be the conformal blocks of a nontrivial,
interacting conformally invariant theory. For the case of $AdS_5 \times S^5$ the
$Z^\beta$ are the partition functions of the $SU(N)/\Zh_N$, ${\cal N} = 4$ SYM theory
in the 't Hooft sector $\beta$. The wavefunctions $Z^\beta$ and $Z^{\rm singleton}_\beta$
transform contragrediently under $SL(2,\Zh)$ invariance of the IIB supergravity, reflecting
the $SL(2,\Zh)$ invariance of the dual  $U(N) = (SU(N)\times U(1))/\Zh_N$
${\cal N}=4$ SYM  theory.\footnote{At first sight there is an apparent contradiction
with the existence of a baryon vertex.
These puzzles, and their resolutions are discussed in
\cite{Witten:1998wy,Aharony:1998qu},
\cite{Aharony:1999ti} p.58, \cite{Maldacena:2001ss} appendix B.}
One should note that {\it any} compactification holographically dual to a conformal field
theory should have a partition function of the form \eqref{decomp}. For example, in
the background $AdS_5 \times T^{1,1}$ discussed in
\cite{Klebanov:1998hh} the full gauge group
will be
\begin{equation}
{SU(N) \times SU(N) \times U(1)\over \Zh_N }
\end{equation}
with the $\Zh_N$ diagonally embedded.
Similarly in other generalizations
such as those discussed in
\cite{Gauntlett:2004zh,Martelli:2004wu,Bertolini:2004xf,
Gauntlett:2004yd} the gauge group behaves analogously and is
\begin{equation*}
\frac{SU(N)^k\times U(1)}{\Zh_N}
\end{equation*}
with the $\Zh_N$ diagonally embedded.
Since the $U(1)$ degree of freedom has its origin in
the overall center-of-mass degree of freedom in the D-brane picture, constructions such as
the warped deformed conifold
\cite{Klebanov:2000nc,Klebanov:2000hb,Herzog:2002ih} which add fractional
branes will not have such a singleton sector. The reason is that fractional branes
are pinned at the orbifold point
\cite{Douglas:1996sw}. This is in accord with the fact
that in such geometries the factor of $N$ in \eqref{toplag} is logarithmically running, and
the topological sector only makes sense for $N$ integral.

The methods used in this paper follow those used in \cite{Gukov:2004id}
 in the analogous case of
the AdS${}_3$/CFT${}_2$ correspondence. The same methods can be applied to the
AdS${}_7$/CFT${}_6$ correspondence
to derive the conformal blocks for the M5 brane of M-theory. (In the latter case the
harmonic sector for the $C$-field does {\it not} decouple from the
massive modes, but may be approximated by a free theory
at long distances. The main results
were summarized in \cite{mooreS04}.) The methods of this paper rely on path integrals and
are hence not well adapted to the case where $H^*(M_4,\Zh)$
contains a nontrivial torsion subgroup.
However, as pointed out in \cite{Witten:1998wy}, the theory (including the
second derivative terms)
is naturally formulated in terms of Cheeger-Simons characters.
In appendix~\ref{app:CS} we indicate how our
results appear in this formulation, thus extending our results to the case with torsion.
We also explain there  the tadpole constraint, at the level of integral cohomology.

Finally, it is worth pointing out that the derivation of \eqref{toplag} from the
10-dimensional IIB theory is {\it not} straightforward, contrary to naive expectations.
As is well-known, the
IIB equations of motion do not follow from a Lorentz-covariant action. However, they
can be derived by starting with a Lorentz-covariant action $I_{\text{IIB}}$
(given in equation \eqref{IIBaction} below),
deriving the equations of motion from $\delta I_{\text{IIB}} =0$, and then
imposing the self-duality of
the 5-form on those equations of motion. It is common practice to reduce the action
$I_{\text{IIB}}$ \'{a} la Kaluza-Klein. {\it This procedure can lead to inconsistent theories}, and in particular
leads to \eqref{toplag} with $N$ replaced by $N/2$. Such a normalization would lead to an
inconsistent quantum theory for $N$ odd. The origin of the trouble is that the action $I_{\text{IIB}}$
is not well-defined, because   its Chern-Simons term does not carry a proper normalization.
This does not, of course, imply
any inconsistency in the type IIB supergravity, but it does underscore the fact that the
topological phases in the IIB partition functions are very subtle.

\section{IIB conventions. Phase of IIB on $X_5\times Y_5$}
\label{sec:IIBconv}
\dopage{\finkfile}
\setcounter{equation}{0}
The IIB equations of motion can be derived by starting with
a Lorentz invariant action on a spin manifold $X_{10}$ and then imposing the
self-duality constraint. The action in the
Einstein frame is:
\begin{multline}
e^{iI_{\text{IIB}}}=\exp\biggl[\frac{2\pi i}{g_B^2\ell_s^8}\int_{X_{10}}\sqrt{-g}\,
\Bigl[\mathcal{R}-\frac{1}{2\tau_2^2}\,\nabla_{\mu}\bar{\tau}\nabla^{\mu}\tau
\Bigr]
-\frac{i\pi}{2}\int_{X_{10}}R_5\wedge * R_5
\\
-\frac{i\pi}{g_B\ell_s^4}
\int_{X_{10}}\frac{1}{\tau_2}(R_3+i\tau_2 H_3)\wedge*(R_3-i\tau_2 H_3)
\biggr]\,\Phi_B
\label{IIBaction}
\end{multline}
where $\tau=C_0+i\tau_2$ and $\Phi_B$ are given below.
The Bianchi identities are
\begin{equation}
dR_1 =0,\qquad  dH_3=0,\qquad dR_3-H_3\wedge R_1=0,\qquad
dR_5-H_3\wedge R_3=0.
\label{Bianchi1}
\end{equation}
Locally they can be solved by
\begin{equation}
H_3=dB_2,\qquad R_1=dC_0,\qquad R_3=dC_2-H_3 C_0,\qquad R_5=dC_4-C_2\wedge H_3.
\label{Bsol}
\end{equation}
The phase $\Phi_B$ is very subtle.
Naively this phase is
\begin{equation*}
\Phi_B=\exp\Bigl[i\pi\int_{X_{10}}C_4\wedge H_3\wedge dC_2\Bigr].
\end{equation*}
After obtaining equations of motion by varying the action with respect to
the potentials $B_2,\,C_0,\,C_2,\,C_4$ one  must impose
by hand the  additional constraint $R_5=-*R_5$.
Of course, the equations of motion obtained this way
do not follow from a Lorentz invariant action. If one ignores
this and dimensionally reduces \eqref{IIBaction} anyway,
one can obtain  an inconsistent quantum theory.

Our considerations are rather general, but for definiteness
we note that they apply to   Freund-Rubin type backgrounds.
The space $X_{10}$ is a product $X_5\times Y_5$, where
$Y_5$ is a compact manifold. The metric on $X_{10}$
is a product metric $ds^2=ds^2_{X_5}+R^2ds^2_{Y_5}$. We choose the $5$-form
flux to be
\begin{equation}
R_5=\frac{N}{\mathrm{Vol}(Y_5)}\,\bigl[\vol(Y_5)-*_{10}\vol(Y_5)\bigr].
\label{R5}
\end{equation}
Here $\vol(Y_5)$ is the volume form on $Y_5$, and $\mathrm{Vol}(Y_5)$
is the volume of the compact manifold $Y_5$ (in our conventions
it is dimensionless).
Then all equation can be satisfied if we take $\tau=C_0+i g_B^{-1}=\mathrm{const}$,
$F_3=H_3=0$ and
\begin{equation*}
\mathcal{R}_{\mu\nu}(X_5)=-\frac{4}{R^{2}}\,g_{\mu\nu}(X_5)\quad\text{and}\quad
\mathcal{R}_{IJ}(Y_5)=4\, g_{IJ}(Y_5)
\end{equation*}
where
\begin{equation}
R=\ell_s\left[\frac{g_B N}{4\mathrm{Vol}(Y_5)}\right]^{1/4}.
\label{R}
\end{equation}
One sees that $X_5$ and $Y_5$ are negatively and positively curved Einstein manifolds
respectively.
We will suppose that $X_5$
has a conformal boundary $M_4$.
For $Y_5$ we consider two examples:   $Y_5 = S^5$ and  $Y_5 = T^{1,1}$ \cite{Klebanov:1998hh}.

Now we want to take into account fluctuations of
the fields $B_2$ and $C_2$. To this end we need to know
the phase of the IIB theory. Although it is
not clear how to obtain it directly from the IIB functional \eqref{IIBaction},
one can get it indirectly. Fortunately both
$S^5$ and $T^{1,1}$ can be considered as $S^1$ bundles
over $\Ch P^2$ and $S^2\times S^2$ respectively.
One can compactify the theory on this $S^1$
and do   $T$-duality. This untwists the bundle and adds   $H_3$ flux
into the IIA background.

\subsection{$Y_5=S^5$}
This case was considered in \cite{Duff:1998us}. Any odd dimensional
unit sphere $S^{2n+1}$ can be represented as
an $S^1$ bundle over $\Ch P^n$. Let
$0\leqslant\sigma<1$ be a coordinate on the $S^1$.
The metric
on the unit $S^{2n+1}$ sphere can be written as
\begin{equation*}
d\Omega_{2n+1}^2=(d\sigma+A)^2+ds_{\Ch P^n}^2
\end{equation*}
where $ds^2_{\Ch P^n}$ is the Fubini-Study metric on $\Ch P^n$
and $A$ is the $1$-form on $\Ch P^n$. The Ricci tensor
of the metric $g_{IJ}$ of the unit sphere $S^{2n+1}$ is
$\mathcal{R}_{IJ}(S^{2n+1})
=2n\,g_{IJ}$.
The metric $g_{ij}$ of $\Ch P^n$ is normalized such that
its Ricci tensor is $\mathcal{R}_{ij}(\Ch P^n)=(2n+2)\,g_{ij}$.
We also have to require that the curvature $F$ of
the $U(1)$ gauge field $A$ must equal $2J$ where
$J$ is the K\"{a}hler form on $\Ch P^n$.
The volume form of sphere decomposes as
\begin{equation}
\vol(S^{2n+1})=\vol(\Ch P^n)\wedge d\sigma
\quad\text{where}\quad
\vol(\Ch P^n)=\frac{1}{n!}J^n.
\label{S2n1}
\end{equation}

The phase for the IIB theory on $X_5\times S^5$ can be obtained
as follows. Consider IIB theory on $X_5\times S^5$ where
$S^5$ is represented as the Hopf fibration
$S^1\to S^5\to\Ch P^2$,
\begin{equation*}
ds^2_{10}=ds^2_{X_5}+R^2[ds^2_{\Ch P^2}+(d\sigma+A^{\text{(R)}})^2]
\end{equation*}
where  $R$ is the radius of $S^5$, and $A^{\text{(R)}}$ is a
connection form with curvature $ \bar F_2^{\text{(R)}}=2J$.
 Then perform the $T$-duality
transformation over $S^1$ and obtain
IIA theory on $X_5\times \Ch P^2\times S^1$ with the
nontrivial flux $H_3=\bar{H}_2\wedge(d\sigma+A^{\text{(R)}})$ \cite{Duff:1998us}.
Notice that the $T$-duality untwists the Hopf fibration
and turns it into the direct product. The IIA
phase is well defined because it can be obtained by
the reduction of the $M$-theory phase \cite{Diaconescu:2003bm}.
The $5$-form field strength reduces as $R_5=\bar{R}_5+\bar{R}_4\wedge
(d\sigma+A^{\text{(R)}})$. From Eqs.~\eqref{R5} and \eqref{S2n1}
one finds
\begin{equation}
\bar{R}_4=N\vol(\Ch P^2).
\label{R4}
\end{equation}
Careful matching of the IIA and IIB fluxes on $X_9\times S^1$
shows that the IIB phase on $X_5\times S^5$ is
\begin{equation}
\Phi_B=\exp\Bigl[-i\pi\int_{Z_6\times \Ch P^2}(\bar{R}_4^2F_2^{\text{(R)}}
+2\bar{R}_4\bar{R}_3\bar{H}_3)\Bigr]
=\exp\Bigl[-2\pi i N\int_{Z_6}\bar{R}_3\wedge\bar{H}_3\Bigr]
\label{PB}
\end{equation}
where $\pd Z_6=X_5$, and we use Eq.~\eqref{R4}
to obtain the last equality. $\bar{R}_3$ and $\bar{H}_3$
comes from the reduction of $R_3$ and $H_3$ to $X_5$.
Notice the ``extra'' factor of $2$ in front of the integral.
This justifies \eqref{toplag}.

\subsection{$Y_5=T^{1,1}$}

$T^{1,1}$ can be considered as an $S^1$ bundle over $S^2\times S^2$.
The metric is \cite{Herzog:2002ih}
\begin{equation}
ds^2_{T^{1,1}}=\frac{1}{9}\,(d\psi+4\pi A^{\text{(R)}})^2+\frac{1}{6}\Bigl[
d\theta^2_{i}+\sin^2\theta_i\,d\phi_i^2
\Bigr],\quad
A^{\text{(R)}}=\frac{1}{4\pi}\bigl[\cos\theta_1 d\phi_1+\cos\theta_2 d\phi_2\bigr]
\label{T11m}
\end{equation}
where $i=1,2$, $0\leqslant\theta_i\leqslant \pi$ and $0\leqslant \phi_i <2\pi$
are coordinates on two $S^2$, $0\leqslant \psi <4\pi$
is coordinate on $S^1$ and $A^{\text{(R)}}$ is a connection.
The curvature of this connection is
\begin{equation*}
F=-\omega_1-\omega_2\quad\text{where}\quad
\omega_i=\frac{1}{4\pi}\,\sin\theta_i\,d\theta_i\wedge d\phi.
\end{equation*}
Here $\omega_i$ is a generator of $H^2(S^2,\Zh)$.
The metric \eqref{T11m} is normalized such that
$\mathcal{R}_{IJ}(T^{1,1})=4g_{IJ}(T^{1,1})$.
Consider IIB theory on $X_5\times T^{1,1}$:
\begin{equation*}
ds^2_{10}=ds^2_{X_5}+R^2\,ds^2_{T^{1,1}}
\end{equation*}
where $R$ is a ``radius'' of $T^{1,1}$.
 Then perform the $T$-duality
transformation over $S^1$ and obtain
IIA theory on $X_5\times S^2\times S^2\times S^1$ with the
nontrivial flux $H_3=\bar{H}_2\wedge(\frac{1}{4\pi}d\psi+A^{\text{(R)}})$.
Notice that the $T$-duality untwists the fibration
and turns it into the direct product.
The $5$-form field strength reduces as $R_5=\bar{R}_5+\bar{R}_4\wedge
(\frac{1}{4\pi}d\psi+A^{\text{(R)}})$. From Eqs.~\eqref{R5} and \eqref{S2n1}
one finds
\begin{equation}
\bar{R}_4=N\,\omega_1\wedge\omega_2.
\label{R4T}
\end{equation}
Careful matching of the IIA and IIB fluxes on $X_9\times S^1$
shows that the IIB phase on $X_5\times S^2\times S^2\times S^1$ is
\begin{equation}
\Phi_B=\exp\Bigl[-i\pi\int_{Z_6\times S^2\times S^2}(\bar{R}_4^2F_2^{\text{(R)}}
+2\bar{R}_4\bar{R}_3\bar{H}_3)\Bigr]
=\exp\Bigl[-2\pi i N\int_{Z_6}\bar{R}_3\wedge\bar{H}_3\Bigr]
\label{PBT}
\end{equation}
where $\pd Z_6=X_5$, and we use Eq.~\eqref{R4T}
to obtain the last equality. $\bar{R}_3$ and $\bar{H}_3$
comes from the reduction $R_3$ and $H_3$ on $X_5$.

Notice that the phase \eqref{PBT} of IIB on $X_5\times T^{1,1}$
and the phase \eqref{PB} of IIB on $X_5\times S^5$ is the same.
In this way  we arrive at the topological term \eqref{toplag}.

\subsection{$5$D Lagrangian for $BC$ fields}
The $BC$ part of the Kaluza-Klein reduction of IIB on $X_5\times Y_5$ is
\begin{subequations}
\begin{equation}
e^{iS_{BC}}
=\exp\biggl[-\frac{i \nu}{2}\int_{X_5} \,
\begin{pmatrix}
F_3 & H_3
\end{pmatrix}
\Mc(\tau)
\begin{pmatrix}
*F_3
\\
*H_3
\end{pmatrix}
\biggr]
\,\Phi_B(B_2,C_2),
\label{iib}
\end{equation}
where $\nu=4\pi R^5\mathrm{Vol}(Y_5)/g_B^2\ell_s^8$ and $R$ is given
in \eqref{R}, \textit{locally} $F_3=dC_2,\,H_3=dB_2$,
$\tau=\text{const}$
and $\Mc(\tau)$ is correspondingly the complex structure and the metric on the torus
\begin{equation}
\Mc(\tau)=\frac{1}{\Im\tau}
\begin{pmatrix}
1 & -\tau_1
\\
-\tau_1 & |\tau|^2
\end{pmatrix},\qquad\det\Mc(\tau)=1.
\label{M(tau)}
\end{equation}
$*$ is Hodge dual with respect to the metric on $X_5$.

The phase $\Phi_B$ is defined by
\begin{equation}
\Phi_B(B_2,\,C_2)=\exp\Bigl[2\pi i N\int_{Z_6}H_3\wedge F_3\Bigr].
\label{phiB}
\end{equation}
\end{subequations}
While $B_2,C_2$ need not be globally well-defined, their
fieldstrengths are well-defined. In this formula we have
extended $H_3,F_3$ to a bounding $6$-fold.
\footnote{The expression is more properly defined in terms of
Cheeger-Simons characters, as indicated in appendix A.}
Suppose we shift $B_2 \to B_2 + b_2$,
$C_2 \to C_2 + c_2$, where $b_2,c_2$ are globally well-defined on $X_5$.
In this case we have the variational formula:
\begin{equation}
\Phi_B(B_2+b_2, C_2 + c_2) = \Phi_B(B_2,C_2)
\exp\biggl[2\pi i N\int_{X_5} \bigl(b_2 F_3 - c_2 H_3)  + i\pi N\int_{X_5}\bigl(b_2\wedge dc_2-c_2\wedge d b_2\bigr)
\biggr]
\label{phib1}
\end{equation}
When $X_5$ has a nonzero boundary then $\Phi_B$ must be considered as
a section of a line bundle. In writing the last factor of \eqref{phib1}
we have chosen a trivialization which is well-adapted to showing the
$SL(2,\Zh)$ invariance. Other choices differ by a total derivative.

The action \eqref{iib} is invariant under   $SL(2,\Zh)$
transformations. This duality group acts on the fields as follows
\begin{equation*}
\Lambda=\begin{pmatrix}
a & b
\\
c & d
\end{pmatrix},\quad ad-bc=1;\qquad
\begin{pmatrix}
F_3'
\\
H_3'
\end{pmatrix}
=\Lambda\begin{pmatrix}
F_3
\\
H_3
\end{pmatrix},\qquad
\tau'=\frac{a\tau+b}{c\tau+d}.
\end{equation*}

The classical equations of motion for $F_3$ and $H_3$ are
\begin{equation}
\nu\,d\biggl[\Mc(\tau)
\begin{pmatrix}
*F_3
\\
*H_3
\end{pmatrix}
\biggr]+2\pi N
\begin{pmatrix}
-H_3
\\
F_3
\end{pmatrix}
=0.
\label{eom}
\end{equation}
This equation implies that $F_3$ and $H_3$ are
trivial in   cohomology.

Near the conformal boundary the manifold $X_5$ looks like
a product $\Rh_+\times M_4$. Further we will assume
that $M_4$ is a compact manifold.
We are working in a Euclidean formulation of AdS/CFT.

The metric
in the vicinity of the conformal boundary
is $ds^2_{X_5}=dr^2/r^2+r^{-2}ds^2_{M_4}$. The slice $r=0$
corresponds to the conformal boundary $M_4$.
More generally we consider
metrics of the form
\begin{equation}
ds^2_{X_5}=d\rho^2+\Omega^2(\rho)\,ds_{M_4}^2
\label{metric}
\end{equation}
where $\rho\in\Rh$. The conformal boundary is located at $\rho=+\infty$.
The orientation is $d\rho\wedge d^4x$.
We will consider $\rho$ to be a Euclidean time variable $\rho=-it$,
and work out the Hamiltonian formalism.

Consider now reduction of the field $F_3$
(the discussion for $H_3$ is similar).
It reduces as
\begin{equation*}
F_3=\bar{F}(t)+dt\wedge \bar{F}_0(t)
\end{equation*}
where $\bar{F}$ and
$\bar{F}_0$ are $3$- and $2$-forms on $M_4$
respectively.
The Bianchi identities are
\begin{equation}
d\bar{F}=0,\quad \pd_{0}\bar{F}-d\bar{F}_0=0
\label{Bianci2}
\end{equation}
where $d$ is differentiation along $M_4$.

At this point we use the Gauss law to conclude that
 $F_3$ is topologically trivial, so
the global solution of these Bianchi identities is
\begin{equation}
\bar{F}(t)=d\bar{c}(t)\quad\text{and}
\quad
\bar{F}_0(t)=\pd_{0}\bar{c}(t)-d\bar{c}_0(t)
\label{redFH}
\end{equation}
where $\bar c(t), \bar c_0(t)$ are globally well-defined.

Now we want to rewrite the action \eqref{iib} in the new variables.
 Substituting \eqref{redFH} into \eqref{phib1}
one
can write the action as $S_{BC}=\int dt(\mathscr{L}_1+\mathscr{L}_2)$
where
\begin{subequations}
\begin{align}
\mathscr{L}_1&=\frac{\nu}{2}\int_{M_4}
\begin{pmatrix}
\bar{F}_0 & \bar{H}_0
\end{pmatrix}
\Mc(\tau)
\begin{pmatrix}
*_4\bar{F}_0
\\
*_4\bar{H}_0
\end{pmatrix}
+\pi N \int_{M_4}\bigl(b_0 \wedge d\bar{c}-c_0\wedge d\bar{b}\bigr)
+\bigl(\bar{b} \wedge \bar{F}_0
-\bar{c}\wedge \bar{H}_0\bigr);
\\
\mathscr{L}_2&=-\frac{\nu}{2\Omega^{2}}
\int_{M_4}
\begin{pmatrix}
\bar{F} & \bar{H}
\end{pmatrix}
\Mc(\tau)*_4
\begin{pmatrix}
\bar{F} \\
\bar{H}
\end{pmatrix}
;
\end{align}
\label{lagr2E}
\end{subequations}
and $\bar{F},\,\bar{H}$ and $\bar{F}_0,\,\bar{H}_0$ are defined in \eqref{redFH}.

\subsection{The Momentum}
The momenta are defined by
\begin{equation*}
\delta S_{BC}=\int_{\Rh}dt\,\int_{M_4}\vol(g)\,\Bigl[
\frac{1}{2}\,\pi_{\bar{c}}^{ij}\,\delta(\pd_0\bar{c}_{ij})
+\frac{1}{2}\,\pi_{\bar{b}}^{ij}\,\delta(\pd_0\bar{b}_{ij})+\dots\Bigr],
\end{equation*}
where ``$\dots$'' denotes variation of the other fields.
Geometrically the momentum is a skewsymmetric
bivector field on $M_4$. However it is more convenient
to regard the momentum as a $2$-form on $M_4$ and define
\begin{equation}
\delta S_{BC}=\int_{\Rh}dt\,\int_{M_4}\Bigl[
\Pi_{\bar{c}}\wedge\delta(\pd_0\bar{c})
+\Pi_{\bar{b}}\wedge\delta(\pd_0\bar{b})+\dots\Bigr].
\end{equation}
The relation between these two definitions is
$\sqrt{g}\,\pi^{ij}_{\bar{c}}=\frac{1}{2}\,\varepsilon^{klij}(\Pi_{\bar{c}})_{kl}$.
In our conventions $\varepsilon^{klij}\in\{\pm 1,0\}$
and $\varepsilon^{1234}=+1$.

Using \eqref{lagr2E} it is straightforward to show that
\begin{equation}
\begin{pmatrix}
\Pi_{\bar{c}}
\\
\Pi_{\bar{b}}
\end{pmatrix}
=
\begin{pmatrix}
\tilde{\Pi}_{\bar{c}}
+\pi N \bar{b}
\\
\tilde{\Pi}_{\bar{b}}
-\pi N\bar{c}
\end{pmatrix}
\quad\text{and}\quad
\begin{pmatrix}
\tilde{\Pi}_{\bar{c}}
\\
\tilde{\Pi}_{\bar{b}}
\end{pmatrix}
=\nu\,\Mc(\tau)
*_4\!
\begin{pmatrix}
\bar{F}_0
\\
\bar{H}_0
\end{pmatrix}.
\label{momentum}
\end{equation}
Here $\bar{F}_0,\,\bar{H}_0$ are defined in \eqref{redFH}.
The symplectic form $\Omega$ is
\begin{equation}
\Omega=\int_{M_4}\,\delta\Pi_{\bar{c}}\wedge\delta\bar{c}
+\delta\Pi_{\bar{b}}\wedge\delta\bar{b}.
\label{Omega}
\end{equation}

\subsection{The Hamiltonian}
\label{sec:hamiltonian}
The Hamiltonian $\mathscr{H}=\mathscr{H}_e+\mathscr{H}_m$ is given by the Legendre transform
\begin{equation}
\mathscr{L}=\int_{M_4}
\bigl[\Pi_{\bar{c}}\wedge(\pd_0\bar{c}-d\bar{c}_0)
+\Pi_{\bar{b}}\wedge(\pd_0\bar{b}-d\bar{b}_0)\bigr]-\mathscr{H}_e
-\mathscr{H}_m
+\pi N\int_{M_4}\bar{b}_0\wedge d\bar{c}-\bar{c}_0\wedge d\bar{b}.
\label{lagr}
\end{equation}
Straightforward calculation shows that
\begin{subequations}
\begin{align}
\mathscr{H}_m&=
\frac{\nu}{2\,\Omega^{2}}\int_{M_4}
\frac{1}{\tau_2}\,
F_{\boldsymbol{\tau}}
*_4F_{\boldsymbol{\bar{\tau}}}
;
\label{Hm}
\\
\mathscr{H}_e&=
\frac{1}{2\nu\tau_2}\int_{M_4}
\bigl[
\tilde{\Pi}_{\boldsymbol{\tau}}
*_4\tilde{\Pi}_{\boldsymbol{\bar{\tau}}}
+
\tilde{\Pi}_{\boldsymbol{\bar{\tau}}}
*_4\tilde{\Pi}_{\boldsymbol{\tau}}
\bigr]
\label{He}
\end{align}
\end{subequations}
where
\begin{equation}
\tilde{\Pi}_{\boldsymbol{\tau}}=\tilde{\Pi}_{\bar{b}}+\tau\,\tilde{\Pi}_{\bar{c}}
\quad\text{and}\quad
\tilde{\Pi}_{\boldsymbol{\bar{\tau}}}=\tilde{\Pi}_{\bar{b}}+\bar{\tau}\,\tilde{\Pi}_{\bar{c}}
\label{Pitau}
\end{equation}
and $F_{\boldsymbol{\tau}}=d\bar{c}-\tau\,d\bar{b}$.
The duality group $SL(2,\Zh)$ acts as follows
\begin{equation}
\begin{split}
\Lambda=\begin{pmatrix}
a & b
\\
c & d
\end{pmatrix}:
\qquad
&\tau'=\frac{a\tau+b}{c\tau+d}\quad\text{and}\quad
\begin{pmatrix}
\tilde{\Pi}_{\bar{c}}
\\
\tilde{\Pi}_{\bar{b}}
\end{pmatrix}^{\prime}
=\Lambda^{-1}
\begin{pmatrix}
\tilde{\Pi}_{\bar{c}}
\\
\tilde{\Pi}_{\bar{b}}
\end{pmatrix},\quad
\begin{pmatrix}
\bar{F}
\\
\bar{H}
\end{pmatrix}^{\prime}
=\Lambda
\begin{pmatrix}
\bar{F}
\\
\bar{H}
\end{pmatrix};
\\
&\tilde{\Pi}_{\boldsymbol{\tau}'}'=(c\tau+d)^{-1}\tilde{\Pi}_{\boldsymbol{\tau}}
\quad\text{and}\quad
F_{\boldsymbol{\tau}'}'=(c\tau+d)^{-1} F_{\boldsymbol{\tau}}.
\end{split}
\label{Lambda}
\end{equation}

\section{Gauge group. Classical and Quantum Gauss Laws}
\dopage{\finkfile}
\setcounter{equation}{0}
The original $5$-dimensional action \eqref{iib} is invariant
under the following gauge transformations
\begin{equation*}
C_2\mapsto C_2+\omega_C,\qquad B_2\mapsto B_2+\omega_B
\end{equation*}
where $\omega_C$ and $\omega_B$ are closed $2$-forms
with integral periods
on $X_5$. Reduction of these gauge transformations
to $\Rh\times M_4$ yields
\begin{equation}
\begin{split}
\bar{c}\mapsto\bar{c}+\omega_{c},\quad
\bar{c}_0\mapsto\bar{c}_0+\lambda_c
\quad\text{where}\quad
d\omega_{c}=0,\quad \pd_0\omega_c-d\lambda_c=0;
\\
\bar{b}\mapsto\bar{b}+\omega_{b},\quad
\bar{b}_0\mapsto\bar{b}_0+\lambda_b
\quad\text{where}\quad
d\omega_{b}=0,\quad \pd_0\omega_b-d\lambda_b=0.
\end{split}
\end{equation}
If $\omega_c=d\alpha_c$ then the gauge transformation is
called small, while if $\omega_c$ represents some
nontrivial cohomology class in $H^2(M_4,\Zh)$
the gauge transformation is called large.
\footnote{A more careful formulation of the $B,C$ fields
shows that the underlying gauge invariance is more subtle,
but we will not need this level of depth for the present
paper. See appendix A for an indication of what is involved. }

It is easy to see that the momenta $\tilde{\Pi}_{\bar{c}}$
and $\tilde{\Pi}_{\bar{b}}$ are gauge invariant.
The parts $\mathscr{H}_e$ and $\mathscr{H}_m$
of the Hamiltonian are separately invariant under
the gauge transformations.

\subsection{Classical Gauss law. Naive quantization}
The part of the Lagrangian \eqref{lagr2E} containing
the Lagrangian multipliers $\bar{b}_0$ and $\bar{c}_0$ is
\begin{equation*}
\mathcal{L}_0=
-\int_{M_4}\bigl[\Pi_{\bar{c}}\wedge
d \bar{c}_0+ \Pi_{\bar{b}}\wedge d \bar{b}_0
\bigr]
+\pi N\int_{M_4}
(\bar{b}_0\wedge d\bar{c}-\bar{c}_0\wedge d\bar{b}).
\end{equation*}
The variation with respect $b_0$ and $c_0$
yields the classical Gauss law
\begin{equation}
\mathcal{G}_c
\equiv\frac{\delta\mathscr{L}_0}{\delta \bar{c}_0}
=-d\,(\Pi_{\bar{c}}+\pi N\bar{b})=0
\quad\text{and}\quad
\mathcal{G}_b\equiv
\frac{\delta\mathscr{L}_0}{\delta \bar{b}_0}
=d\,(-\Pi_{\bar{b}}+\pi N\bar{c})=0.
\label{GaussLAW}
\end{equation}
In the quantum theory the
small gauge transformations are generated by this Gauss law.
If $(\alpha_c,\,\alpha_b)$ is a pair of
$1$-forms then the Gauss law for the small transformations
becomes the constraint
\begin{equation}
\Psi(\bar{c},\bar{b})=
e^{-i\int_{M_4}\alpha_c\wedge \mathcal{G}_c
+\alpha_b\wedge\mathcal{G}_b}\,\Psi(\bar{c},\bar{b})
=e^{i\pi N\int_{M_4}
d\alpha_c\wedge \bar{b}-
d\alpha_b\wedge \bar{c}
}\,\Psi(\bar{c}+d\alpha_c,\,\bar{b}+d\alpha_b).
\label{ttact}
\end{equation}
on gauge invariant wavefunctions.

How shall we generalize this to the large gauge transformations?
The natural guess is to replace $d\alpha$ by $\omega$,
a closed $2$-form with integral periods. We define
\begin{equation*}
U(\omega_c,\omega_b)=\exp\Bigl[2\pi i\int_{M_4}
\omega_c\wedge P_{c}+\omega_b\wedge P_b\Bigr]
\end{equation*}
where
\begin{equation}
P_c=\frac{1}{2\pi}\,\Pi_{\bar{c}}+\frac{N}{2}\bar{b}\quad\text{and}\quad
P_b=\frac{1}{2\pi}\,\Pi_{\bar{b}}-\frac{N}{2}\bar{c},
\end{equation}
are conserved ``Page charges.''
Simple calculation shows that these charges do not commute
\begin{equation}
\biggl[\int_{M_4}\omega_c\wedge P_c,\,\int_{M_4}\omega_b\wedge P_b\biggr]=
\frac{i N}{2\pi}\int_{M_4}
\omega_c\wedge \omega_b.
\label{PP}
\end{equation}

The naive calculation,  performed in the same way as the
previous one, suggests
\begin{equation*}
\bigl(U(\omega_{c},\omega_{b})\Psi\bigr)(\bar{c},\bar{b})=
\exp\Bigl[+i\pi N\int_{M_4}\omega_{c}\wedge \bar{b}-\omega_{b}\wedge \bar{c}
\Bigr]\,
\Psi(\bar{c}+\omega_{c},\bar{b}+\omega_{b}).
\label{UPsi}
\end{equation*}
Applying $U$ twice one obtains the following group law
\begin{equation}
U(1)U(2)=\exp\Bigl\{i\pi N\int_{M_4}
\omega^{(2)}_c\wedge\omega^{(1)}_b-\omega_b^{(2)}\wedge\omega_c^{(1)}\Bigr\}\,U(1+2).
\label{gausscocycle}
\end{equation}
This indicates that there is a $\Zh_2$ anomaly in the quantum Gauss law, suggesting
the theory is not consistent for $N$ odd. In fact, the theory {\it is} consistent for all
$N$, and the above procedure is simply too naive, as it ignores the key geometrical
fact that the wavefunction must be considered as a section of a line bundle with
nonzero curvature. We will explain this in the next subsection.

The operators $P_c,P_b$ are not gauge invariant. Following
the discussion in \cite{mooreS04} we define
gauge invariant quantities
\begin{subequations}
\begin{equation}
W_c(\phi)=e^{2\pi i\int_{M_4}\phi\wedge P_c}
\quad\text{and}\quad
W_b(\phi)=e^{2\pi i\int_{M_4}\phi\wedge P_b}
\end{equation}
where $\phi$ is an arbitrary $2$-form. These operators
satisfy the following commutation relation:
\begin{equation}
W_c(\phi_1)W_b(\phi_2)=e^{-2\pi i N\int_{M_4}\phi_1\wedge\phi_2}\,
W_b(\phi_2)W_c(\phi_1).
\end{equation}
\label{Hgroup}
\end{subequations}
Under gauge transformations these operators
change as follows:
\begin{equation}
\begin{split}
U(\omega_c,\omega_b)W_c(\phi) U^{-1}(\omega_c,\omega_b)
&=W_c(\phi)\,e^{+2\pi iN\int_{M_4}\phi\wedge \omega_b};
\\
U(\omega_c,\omega_b)W_b(\phi) U^{-1}(\omega_c,\omega_b)
&=W_b(\phi)\,e^{-2\pi iN\int_{M_4}\phi\wedge \omega_c};
.
\end{split}
\end{equation}
One sees that $W_c(\phi)$ and $W_b(\phi)$ are gauge invariant
if $N\phi$ is in $\Omega^2_{\Zh}(M_4)$. Notice that
if $\phi$ is a closed $2$-form with integral periods then the
corresponding $W$'s are just gauge transformations.
Therefore we can identify $\phi$ with $\phi+\xi$
where $\xi$ is any vector in $H^2(M_4,\Zh)$.
With this identification $W_b$ and $W_c$ generate  the
finite Heisenberg group $W$:
\begin{equation*}
0\to\Zh_N\to W\to H^2(M_4,\Zh_N)\times H^2(M_4,\Zh_N)\to 0.
\end{equation*}
The Hilbert space of the theory should be a representation of
this group, and we will confirm this below.
$W$ is the magnetic translation group analogous
to that of the $M$-theory $3$-form in \cite{mooreS04}.
{}From the dual gauge theory point of view
$W$ is related to the 't Hooft lattice of the
discrete electric and magnetic charges.

\subsection{Quantum Gauss Law}

In equation \eqref{gausscocycle} we found a potential anomaly in the
Gauss law. The resolution of this problem lies in the fact that the
wavefunction must be considered as a section of a line bundle $\mathcal{L}_N$
over the space of gauge-invariant field configurations satisfying the
Gauss law. \footnote{The following discussion
is closely related to section~6 in \cite{Diaconescu:2003bm}.}
In this section we assume $H^*(M_4,\Zh)$ is torsion free. The
generalization to the case with torsion is indicated in appendix A. Thus,
in this section,
the Gauss law shows that $B_2,C_2$ are globally well-defined

%

The wave function is a section of a line bundle $\mathcal{L}_N$
over the space of pairs of $2$-forms $\Omega^2(M_4) \times \Omega^2(M_4)$.
This line bundle has a natural connection defined by the phase \eqref{phib1}.
Consider path $p(t)=(C(t),B(t))$ in the space of  forms,
$t\in[0,1]$ is the coordinate on the path. Then the parallel
transport is defined by  \eqref{phib1}:
\begin{equation}
U(p) =\Phi_B(C(t),B(t))\in \mathrm{Hom}
\bigl(\mathcal{L}_N\bigl|_{(C(0),B(0))}, \mathcal{L}_N\bigl|_{(C(1),B(1))}\bigr)
\label{connection}
\end{equation}
It is straightforward to compute the curvature
of \eqref{connection}
\begin{equation}
\Omega\bigl((\phi_c^{(1)},\phi_b^{(1)}),\,
(\phi_c^{(2)},\phi_b^{(2)})\bigr)
=2\pi i N\int_{M_4}\bigl(\phi^{(1)}_{c}\wedge\phi^{(2)}_b
-\phi^{(1)}_b\wedge \phi^{(2)}_{c}\bigr)
\label{Omegacurv}
\end{equation}
where  $\phi_b^{(i)} ,\,\phi_c^{(i)}$ are arbitrary $2$-forms.
Now, for any 2-forms $\phi_b,\phi_c$  introduce the straightline path
\begin{equation}
p_{\bar{c},\bar{b};\,\phi_c,\phi_b}(t)=
\{C(t)=\bar{c}+t\phi_c,\;
B(t)=\bar{b}+t\phi_b\}
\label{path}
\end{equation}
Using the formula for the curvature   we find
\begin{equation}
U(p_{\bar{c},\bar{b};\,\phi^{(1)}+\phi^{(2)}})
=
U(p_{\bar{c}+\phi^{(1)}_c,\bar{b}+\phi^{(1)}_b;\,\phi^{(2)}})
\,U(p_{\bar{c},\bar{b};\,\phi^{(1)}})
\exp\Bigl\{i\pi N\int_{M_4}
\bigl(\phi^{(1)}_{c}\wedge\phi^{(2)}_b-\phi^{(1)}_b\wedge\phi^{(2)}_c\bigr)
\Bigr\}.
\label{Umult}
\end{equation}

It follows from \eqref{Umult} that  parallel transport
\textit{does not} define a lift
of the gauge group to the total space of $\mathcal{L}_N$. To define the lift of the
group action we choose the standard path, say \eqref{path}.
Then define the action on a section $\Psi$ of $\mathcal{L}_N$ by
\begin{equation}
\bigl(g(\omega_c,\omega_b)\cdot\Psi\bigr)(\bar{c}+\omega_c,\bar{b}+\omega_b)=
\varphi(\bar{c},\bar{b};\,\omega_c,\omega_b)^*\,
U(p_{\bar{c},\bar{b};\,\omega_c,\omega_b})\,
\Psi(\bar{c},\bar{b})
\label{gact1}
\end{equation}
where $\varphi$ is a phase, and $\omega_c,\,\omega_b$
are closed $2$-forms with integral periods. The  ``lifting phase''
 $\varphi$ must satisfy
\begin{equation}
\varphi(\bar{c},\bar{b};\omega^{(1)}+\omega^{(2)})
=\varphi(\bar{c}+\omega^{(1)}_c,\bar{b}+\omega^{(1)}_b;\omega^{(2)})
\varphi(\bar{c},\bar{b};\omega^{(1)})
\,e^{i\pi N\int_{M_4}
\omega^{(1)}_{c}\wedge\omega^{(2)}_b-\omega^{(1)}_b\wedge\omega^{(2)}_c}.
\label{liftingrel}
\end{equation}

Since we are working in the case where $H^*(M_4,\Zh)$ is torsion free
the lifting phase can be written in terms of local integrals.
The most general solution of \eqref{liftingrel}
 satisfying $\varphi(\bar{c},\bar{b};0,0)=1$ is:
\begin{equation*}
\varphi(\bar{c},\bar{b};\omega_{c},\omega_{b})=
\exp\left\{i\pi\rho\int_{M_4}\omega_{c}\wedge\omega_{b}
+i\pi\alpha\int_{M_4}\bar{b}\wedge\omega_{c}
-i\pi\beta\int_{M_4}\bar{c}\wedge\omega_{b}
\right\}
\end{equation*}
where  $\beta\equiv N-\rho\mod 2\Zh$ and $\alpha\equiv N+\rho\mod 2\Zh$.

We now  trivialize the bundle $\mathcal{L}_N$ by using parallel transport
along the paths \eqref{path} to define a canonical nowhere vanishing
section $S(\bar c, \bar b)$.
The {\it ratio} $\psi(\bar c, \bar b) :=  \Psi(\bar c,\bar b)/S(\bar c ,\bar b)$
is   a function, rather than a section. The action of the   gauge group on this
function is
\begin{equation*}
\bigl(\mathfrak{g}(-\omega_c,-\omega_b)\cdot\psi\bigr)(\bar{c},\bar{b})
=\varphi^*(-\omega_c,-\omega_b;c+\omega_c,b+\omega_b)
\exp\Bigl[-i\pi N\int_{M_4}\bar{b}\wedge\omega_{c}
-\bar{c}\wedge\omega_b
\Bigr]\psi(\bar{c}+\omega_c,
\bar{b}+\omega_b).
\end{equation*}
This action of the gauge group must agree with \eqref{ttact}
for $\omega_c=d\alpha_c$, $\omega_b=d\alpha_b$, therefore one concludes
that $\alpha=\beta=2N$, and
\begin{equation}
\varphi(\bar{c},\bar{b};\omega_c,\omega_b)
=e^{-i\pi N\int_{M_4}\omega_c\wedge\omega_b
+2\pi i N\int_{M_4}\bar{b}\wedge\omega_{c}
-\bar{c}\wedge\omega_{b}}.
\label{cocycle}
\end{equation}
Thus  the gauge transformations are given by
\begin{subequations}
\begin{equation}
(\mathfrak{g}(-\omega_c,-\omega_b)\cdot\psi)(c,b)
=e^*_{\omega_c,\omega_b}(c,b)\psi(c+\omega_c,b+\omega_b)
\end{equation}
where
\begin{equation}
e_{\omega_c,\omega_b}(c,b)
=\exp\Bigl[-i\pi N \int_{M_4} \omega_{b}\wedge
\omega_{c}-i\pi N\int_{M_4}\bar{b}\wedge\omega_{c}
-\bar{c}\wedge\omega_b
\Bigr]
\end{equation}
\label{graction}
\end{subequations}
The Gauss law $g\cdot\Psi(c,b)=\Psi(g\cdot(c,b))$ takes the following form
\begin{equation}
\psi^{\text{phys}}(\bar{c}+\omega_c,
\bar{b}+\omega_b)=e_{\omega_c,\omega_b}(c,b)\psi^{\text{phys}}(\bar{c},\bar{b}).
\label{Gausslaw}
\end{equation}

Under an $SL(2,\Zh)$ transformation by $\Lambda$ \eqref{Lambda}
the cocycle \eqref{cocycle} transforms as
\begin{equation}
\varphi(\Lambda\cdot(c,b);\Lambda\cdot(\omega_c,\omega_b))
=\varphi(c,b;\omega_c,\omega_b)\,e^{i\pi N\int_{M_4}ac\omega_c\wedge\omega_c
+bd\omega_b\wedge\omega_b}.
\end{equation}
This appears to break the $SL(2,\Zh)$ invariance.
However if $M_4$ is a spin manifold, then
the index theorem tells us that $\int_{M_4}\omega\wedge\omega\in 2\Zh$
for $\omega\in H^2(M_4,\Zh)$,
and therefore the exponential factor is 1. So we require $M_4$
to be spin manifold. Put differently, if $M_4$ is not spin then
$SL(2,\Zh)$ does not commute with the gauge projection.
%
%
This is in accord with \cite{Witten:1998wy}. In the case that
$M_4$ is not spin we expect that the theory can be modified to
restore $SL(2,\Zh)$ invariance, for reasons described below,
but we leave this for the future.

\section{Spectrum in the harmonic sector}
\dopage{\finkfile}
\setcounter{equation}{0}

Using the Hodge decomposition we can rewrite  $\bar{b},\,\bar{c}$ as
\begin{equation}
\bar{c}=c^h+c'+c''\quad\text{and}\quad
\bar{b}=b^h+b'+b''
\end{equation}
where $c^h,\,b^h\in\mathrm{Harm}^2(M_4,\Rh)$,
$c',\,b'$ are projections on image of $d^{\dag}$,
and $c'',\,b''$ are projections on image of $d$.
The operators $dd^{\dag}$ and $d^{\dag}d$ are separately
self adjoint with respect to the metric $\langle\alpha_p,\,\beta_p\rangle
=\int_{M_4}\alpha_p *\beta_p$.
Moreover they are orthogonal
with respect to this inner product. Therefore the space
of $2$-forms decomposes as
$\Omega^2(M_4)=\mathrm{Harm}^2(M_4)\oplus \mathrm{im}\,(d d^{\dag},\Omega^2)
\oplus \mathrm{im}\,(d^{\dag}d,\Omega^2)$.

As we will see in a moment there is a factorization of
the Hamiltonian on the harmonic Hamiltonian and the one corresponding
to the massive modes. Therefore the wave function $\Psi$
also factorizes as $\Psi=\Psi_{\text{harm}}\Psi_{\text{massive}}$.
The harmonic Hamiltonian does not depend on time, and hence
we can assume that $\Psi_{\text{harm}}$ is its eigenfunction.
The massive sector has a unique groundstate, but the harmonic
sector has many groundstates, leading to a Hilbert space of
``conformal blocks.'' We will mostly be focussing on the
harmonic sector in what follows.

\subsection{Basis}

We choose a basis $\omega^{\alpha}$ for $\mathrm{Harm}^2_{\Zh}(M_4)$ and
$\omega^n$ for the orthogonal complement.
$\omega^{n'}$ forms a basis for $\mathrm{im}(d^{\dag} d)$,
and $\omega^{n''}$ forms a basis for $\mathrm{im}(dd^{\dag})$.
We can choose $\omega^{n'}$ to be eigenvectors of $d^{\dag}d$,
then $*\omega^{n'}$ are eigenvectors of $d d^{\dag}$.
We define a dual basis by
\begin{equation}
\int_{M_4}\omega^{\alpha}\wedge\hat{\omega}_{\beta}=\delta^{\alpha}{}_{\beta}
\quad\text{and}\quad
\int_{M_4}\omega^{n}\wedge\hat{\omega}_{m}=\delta^{n}{}_{m}.
\end{equation}
So we can expand the fields in this basis
\begin{equation}
\begin{split}
\bar{c}&=c_{\alpha}\omega^{\alpha}+c_n\omega^n,
\qquad
\bar{b}=b_{\alpha}\omega^{\alpha}+b_n\omega^n;
\\
\Pi_{\bar{c}}&=\Pi^{\alpha}_c\hat{\omega}_{\alpha}+
\Pi^n_c\hat{\omega}_n,
\qquad
\Pi_{\bar{b}}=\Pi^{\alpha}_b\hat{\omega}_{\alpha}+
\Pi^n_b\hat{\omega}_n
\end{split}
\label{basisexp}
\end{equation}
We also have metrics
\begin{equation}
h_{\alpha\beta}=\int_{M_4}\hat{\omega}_{\alpha}\wedge*\hat{\omega}_{\beta}
\quad\text{and}\quad
h^{\alpha\beta}=\int_{M_4}\omega^{\alpha}\wedge *\omega^{\beta}
\end{equation}
which are inverse of each other,and the period matrix
$\boldsymbol{\tau}^{\alpha\beta}=\int_{M_4}\omega^{\alpha}\wedge\omega^{\beta}.
$
If
$\{\omega^{\alpha}\}$ is an integral basis
in $H^2(M_4,\Zh)$
then the matrix $\boldsymbol{\tau}^{\alpha\beta}$
has two main properties: $\det\boldsymbol{\tau}=1$, and
both the intersection matrix
$\boldsymbol{\tau}^{\alpha\beta}$ and its inverse $\boldsymbol{\tau}_{\alpha\beta}$
have integer coefficients.

One sees that the dual basis is $\hat{\omega}_{\alpha}
=(\boldsymbol{\tau}^{-1})_{\alpha\beta}\omega^{\beta}$, so
\begin{equation*}
\Pi_c^{\alpha}=\tilde{\Pi}_c^{\alpha}+\pi N b_{\beta}
\boldsymbol{\tau}^{\beta\alpha}\quad\text{and}\quad
\Pi_b^{\alpha}=\tilde{\Pi}_b^{\alpha}-\pi N c_{\beta}
\boldsymbol{\tau}^{\beta\alpha}.
\end{equation*}

The symplectic form \eqref{Omega} becomes
$\Omega=\delta\Pi^{\alpha}_{\bar{c}}\wedge \delta c_{\alpha}
+\delta\Pi^{n}_{\bar{c}}\wedge \delta c_{n}
+\delta\Pi^{\alpha}_{\bar{b}}\wedge \delta b_{\alpha}
+\delta\Pi^{n}_{\bar{b}}\wedge \delta b_{n}
$
and hence
\begin{equation}
[\Pi^{\alpha}_{\bar{c}}(t),\,c_{\beta}(t)]=-i\delta^{\alpha}_{\beta}
\quad\text{and}\quad
[\Pi^{\alpha}_{\bar{b}}(t),\,b_{\beta}(t)]=-i\delta^{\alpha}_{\beta},
\label{Pic}
\end{equation}
and similarly for $\Pi^{n'}$ and $\Pi^{n''}$.

We can choose a basis  of harmonic forms on $M_4$ in which
Hodge $*$-operator acts diagonally:
\begin{subequations}
\begin{align}
*\omega^{\alpha}&=+\omega^{\alpha},\quad \alpha=1,\dots,\mathrm{b}_2^+;
\\
*\omega^{\alpha}&=-\omega^{\alpha},\quad \alpha=\mathrm{b}_2^++1,\dots,\mathrm{b}_2^+
+\mathrm{b}_2^-=\mathrm{b}_2(M_4).
\end{align}
\label{Hodgebasis}
\end{subequations}
Notice that in the basis \eqref{Hodgebasis} the period
matrix is not integral, but is related to the Hodge metric
\begin{equation}
h^{\alpha\beta}=\begin{pmatrix}
\boldsymbol{\tau}_+^{\alpha\beta} & 0
\\
0 & -\boldsymbol{\tau}_-^{\alpha\beta}
\end{pmatrix},
\quad
\boldsymbol{\tau}^{\alpha\beta}=\begin{pmatrix}
\boldsymbol{\tau}_+^{\alpha\beta} & 0
\\
0 & \boldsymbol{\tau}_-^{\alpha\beta}
\end{pmatrix},
\quad \boldsymbol{\tau}_+^{\alpha\beta}>0\quad\text{and}\quad
\boldsymbol{\tau}_-^{\alpha\beta}<0.
\label{bdf}
\end{equation}

The Hamiltonian takes the form
\begin{subequations}
\begin{align}
\mathscr{H}_e&=\frac{1}{4\nu\tau_2}\,
\Bigl[h_{\alpha\beta}\,
\bigl(\tilde{\Pi}_{\boldsymbol{\tau}}^{\alpha}
\tilde{\Pi}_{\boldsymbol{\bar{\tau}}}^{\beta}
+\tilde{\Pi}_{\boldsymbol{\bar{\tau}}}^{\alpha}
\tilde{\Pi}_{\boldsymbol{\tau}}^{\beta}\bigr)
+h_{nm}\,
\bigl(\tilde{\Pi}_{\boldsymbol{\tau}}^{n}
\tilde{\Pi}_{\boldsymbol{\bar{\tau}}}^{m}
+\tilde{\Pi}_{\boldsymbol{\bar{\tau}}}^{n}
\tilde{\Pi}_{\boldsymbol{\tau}}^{m}\bigr)
\Bigr];
\label{Hm1}
\\
\mathscr{H}_m&=\frac{\nu}{2\Omega^2(t)\tau_2}\,
h^{n'm'}\,\lambda_{m'}
F_{\boldsymbol{\tau},n'}
F_{\boldsymbol{\bar{\tau}},m'}
\end{align}
\end{subequations}
where $\lambda_{m'}$ are eigenvalues of $d^{\dag}d$:
$d^{\dag}d\omega_{m}=\lambda_m\omega_m$.
One sees that the wave function factorizes on
the product of the wave function $\psi_h$ depending
on the harmonic modes $(c_{\alpha},\,b_{\alpha})$ and the wave function $\psi_m$
depending on the massive modes $(c_{n'},\,b_{n'})$.
%
%
The harmonic Hamiltonian $\mathscr{H}_{\text{harm}}$ is
defined by the first term in \eqref{Hm1}.

Using the commutation relations \eqref{Pic} one obtains
\begin{equation}
[\tilde{\Pi}^{\alpha}_{\boldsymbol{\tau}},\,\tilde{\Pi}^{\beta}_{\boldsymbol{\tau}}]=0,\quad
[\tilde{\Pi}^{\alpha}_{\boldsymbol{\bar{\tau}}},\,\tilde{\Pi}^{\beta}_{\boldsymbol{\bar{\tau}}}]=0,\quad
[\tilde{\Pi}^{\alpha}_{\boldsymbol{\tau}},\,\tilde{\Pi}^{\beta}_{\boldsymbol{\bar{\tau}}}]=
4\pi N\tau_2\,
\boldsymbol{\tau}^{\alpha\beta}.
\label{crel}
\end{equation}
In the basis \eqref{Hodgebasis} the matrix $\boldsymbol{\tau}^{\alpha\beta}$ has
block diagonal form \eqref{bdf}. Assuming that $N$ is positive one sees that for
$\alpha=1,\dots, \mathrm{b}_2^+$ the operators $\tilde{\Pi}^{\alpha}_{\boldsymbol{\tau}}$ are
annihilation operators, while
for $\alpha=\mathrm{b}_2^++1,\dots, \mathrm{b}_2$
the operators  $\tilde{\Pi}^{\alpha}_{\boldsymbol{\bar{\tau}}}$
are annihilation operators.
The first term in \eqref{Hm1} takes the form
\begin{equation}
\mathscr{H}_{\text{harm}}
\stackrel{\ref{crel}}{=}
\frac{1}{2\nu\tau_2}\left\{
\sum_{\alpha,\beta=1}^{\mathrm{b}_2^+}(\boldsymbol{\tau}_+^{-1})_{\alpha\beta}\,
\tilde{\Pi}^{\alpha}_{\boldsymbol{\bar{\tau}}}\tilde{\Pi}^{\beta}_{\boldsymbol{\tau}}
-
\sum_{\alpha,\beta=\mathrm{b}_2^++1}^{\mathrm{b}_2}(\boldsymbol{\tau}_-^{-1})_{\alpha\beta}\,
\tilde{\Pi}^{\alpha}_{\boldsymbol{\tau}}\tilde{\Pi}^{\beta}_{\boldsymbol{\bar{\tau}}}
\right\}
+\frac{\pi N}{\nu}\mathrm{b}_2(M_4)
\label{Hm2}
\end{equation}
{}From this and the block-diagonal
form of the metric $h_{\alpha\beta}$ it is easy to see that
the ground state function $\Psi_0(b_{\alpha},c_{\alpha})$ must
satisfy
\begin{equation}
\tilde{\Pi}^{\alpha}_{\boldsymbol{\tau}}\Psi_0=0,\quad\alpha=1,\dots,\mathrm{b}_2^+;
\qquad
\tilde{\Pi}^{\alpha}_{\boldsymbol{\bar{\tau}}}\Psi_0=0,\quad\alpha=\mathrm{b}_2^++1,\dots,\mathrm{b}_2.
\label{aa*}
\end{equation}
The ground state energy is $\frac{\pi N}{\nu}\mathrm{b}_2(M_4)$
and depends only on the topology of $M_4$.
The excited states are constructed by acting by the creation operators.
The commutation relations of the momenta and Hamiltonian are
\begin{equation}
[\mathscr{H}_{\text{harm}},\,\tilde{\Pi}^{\alpha}_{\boldsymbol{\tau}}]=
-\frac{2\pi N}{\nu}\mathrm{s}_{\alpha}\tilde{\Pi}^{\alpha}_{\boldsymbol{\tau}}
\quad\text{and}\quad
[\mathscr{H}_{\text{harm}},\,\tilde{\Pi}^{\alpha}_{\boldsymbol{\bar{\tau}}}]=
\frac{2\pi N}{\nu}\mathrm{s}_{\alpha}\tilde{\Pi}^{\alpha}_{\boldsymbol{\bar{\tau}}}
\end{equation}
where $s_{\alpha}=(1_{\mathrm{b}_2^+},-1_{\mathrm{b}_2^-})$.
Therefore the spectrum of this Hamiltonian is equally gapped with
the gap width $2\pi N/\nu$.

The most general solution of equations \eqref{aa*} is
\begin{equation}
\psi_0(c^h,b^h)=\exp\Bigl[
-\frac{\pi N}{2\tau_2}\int_{M_4}(c^h-\tau b^h)*_4 (c^h-\bar{\tau}b^h)
\Bigr]
\,\phi(c_+-\tau b_+,\,c_--\bar{\tau}b_-).
\label{Psi0}
\end{equation}
where $* c_\pm = \pm c_\pm$ and $\phi$ is holomorphic.

We now introduce an overcomplete
basis by choosing $\phi$ to be a linear exponential $\phi_{v_c,v_b}$. Covariance
with respect to $SL(2,\Zh)$ suggests the following choice
\begin{equation}
\phi_{v_c,v_b}(c_+-\tau b_+,\,c_--\bar{\tau}b_-)
=
e^{-\frac{\pi N}{\tau_2}\int_{M_4}(v_c-\bar{\tau}v_b)_+\wedge(c-\tau b)_+
+
\frac{\pi N}{\tau_2}\int_{M_4}(v_c-\tau v_b)_-\wedge(c-\bar{\tau} b)_-
}.
\label{psivcvb}
\end{equation}

\subsection{Averaging over the gauge group}
To obtain the wave function satisfying the Gauss law
\eqref{Gausslaw} it is sufficient to average
the solution \eqref{Psi0} over the large gauge transformations
\eqref{graction}. Hence using $\phi_{v_c,v_b}$
of \eqref{psivcvb} in \eqref{Psi0} the physical wave function is
\begin{equation}
\psi^{\text{phys}}_{v_c,v_b}(c^h,b^h):=
\sum_{\omega_c,\omega_b\in\Lambda}
e_{\omega_c,\omega_b}^*(c^h,b^h)\,\psi_{v_c,v_b}(c^h+\omega_c,b^h+\omega_b)
\label{omegasum}
\end{equation}
where $\Lambda=\mathrm{Harm}^2_{\Zh}(M_4)$ is the lattice
of the harmonic $2$-forms with integral periods.
We now follow a standard procedure and
use Poisson resummation to split this sum in a form so
that we can extract the conformal blocks. The details
are in appendix~\ref{app:GS}.
One finds that the sum \eqref{omegasum}, up to an overall
normalization independent of $b$ and $c$, can be
written as
\begin{equation*}
\Psi^{\text{phys}}_{v_c,v_b}(c,b)=
\sum_{\beta\in\Lambda/\Lambda_N}
\Psi_{\beta}^{\text{phys}}(c,b;\tau)
\Psi_{-\beta}^{\text{phys}}(-v_c,v_b;-\bar{\tau}).
\end{equation*}
where $\Lambda_N\approx \mathrm{Harm}^2_{N\Zh}(M_4)$. The physical
wave function is thereby found to be
\begin{equation}
\Psi_{\beta}^{\text{phys}}(c,b;\tau)=\mathscr{N}(\tau)\,
\Theta_{\Lambda+\frac{1}{N}\beta,\,N/2}(\tau,c,b;*)
\label{psiPhys}
\end{equation}
where $c$ and $b$ are harmonic $2$-forms,
and $\Nc(\tau)$ is a normalization constant which will be fixed later.
The Siegel-Narain $\Theta$-function at level $k$ with
characteristics $c,\,b$ is given by the following series
\begin{equation}
\Theta_{\Lambda+\gamma,\,k}(\tau,c,b;*)
=e^{2\pi ik\int_{M_4}c\wedge b}\sum_{\omega\in\Lambda+\gamma}
e^{2\pi ik\tau\int_{M_4}(\omega+b)_+^2
+2\pi ik\bar{\tau}\int_{M_4}(\omega+b)_-^2
-4\pi i k\int_{M_4}c\wedge(\omega+b)
}
\label{SNth}
\end{equation}
where $\gamma$ is an element of $\frac{1}{2k}\Lambda$.

The magnetic translation group \eqref{Hgroup} acts on the physical wave
functions as follows:
\begin{subequations}
\begin{align}
W_c(\phi)\Psi_{\beta}^{\text{phys}}(c,b;\tau)
&=e^{-2\pi i\int_{M_4}\phi\wedge\beta}\,\Psi_{\beta}^{\text{phys}}(c,b;\tau);
\\
W_b(\phi)\Psi_{\beta}^{\text{phys}}(c,b;\tau)
&=\Psi_{\beta+N\phi}^{\text{phys}}(c,b;\tau).
\end{align}
\label{Wrep}
\end{subequations}
Here it is assumed that $N\phi\in\Lambda$. One sees
that the space of the physical wave functions
is a representation space for $W$. In \cite{Witten:1998wy}
the algebra \eqref{Hgroup} and its representation
\eqref{Wrep} occurs. The description of the
operators in \cite{Witten:1998wy}, (eq. 3.5) compared to our  \eqref{Hgroup}
is different. This happens because we retain the
kinetic terms for the $b,c$ fields.
It is easy to see that in the limit $\nu\to 0$ the operators
$W_c$ and $W_b$ becomes the operators defined in \cite{Witten:1998wy}.
In making this comparison one must regard the cohomology
class $[N\phi]$ as Poincar\'{e} dual to a $2$-cycle in $M_4$.

\subsection{Normalization of the wave function}

Since the Hamiltonian and Hilbert space factorize into flat and massive
sectors we can consider the wavefunction restricted to the flat fields.
We now observe an interesting consequence of normalizing the wavefunction
of the flat fields.

The inner product on the space of flat fields is defined by
\begin{equation}
\langle\Psi_{\beta},\Psi_{\beta'}\rangle:=
\int_{Z^{2}(M_4,\Rh)\times Z^2(M_4,\Rh)}\frac{\Ds_g C\,\Ds_g B}{\mathrm{vol}(\text{gauge group})}\,
\overline{\Psi_{\beta}(B,C)}\,\Psi_{\beta'}(B,C),
\label{inner}
\end{equation}
where the integral runs  over all closed $2$-forms on $M_4$.
The integral descends to one on the space of gauge inequivalent flat fields.
This space is  $\mathbb{T}\times \mathbb{T}$ where
 $\mathbb{T}=H^2_{DR}(M_4,\Rh)/H^2_{DR}(M_4,\Zh)$.
In order to fix the gauge we will follow the recipe
of \cite{Gegenberg:1993gd}.

We use the Hodge decomposition to write
$C=c^h+d\alpha_c$ and $B=b^h+d\alpha_b$, where $c^h$ and
$b^h$ are harmonic $2$-forms, $\alpha_b$ and $\alpha_c$ are $1$-forms.
However $\alpha_b$ and $\alpha_c$ also have   gauge degrees of freedom.
We can fix this gauge freedom by
saying that $\alpha_b=\alpha_b^T$ and $\alpha_c=\alpha_c^T$ are
in the image of $d^{\dag}$.
The measure as usual
can be obtained from the norm:
\begin{equation*}
\|\delta C\|^2_g=\int_{M_4}\delta C *\delta C
=\int_{M_4}\delta c^h*\,\delta c^h
+\int_{M_4}\delta \alpha_c^T *(d^{\dag}d)\delta\alpha_c^T
\quad\Rightarrow\quad
\Ds_g C=\sqrt{\det{}'_1 (d^{\dag}d)}\,\Ds_g c^h\,\Ds_g \alpha_c^T,
\end{equation*}
where $\det{}'_1(d^{\dag}d)$ is the determinant of the
operator $d^{\dag}d$ on the space of $1$-forms, and ${}'$ means that we excluded
the zero modes. Using some
identities \cite{Gegenberg:1993gd} one can rewrite it as the ratio
of the Laplacian operators $\det{}'\Delta_1/\det{}'\Delta_0$.
The integrals over $\Ds_g\alpha_c^T$ and $\Ds_g\alpha_b^T$ partially cancel the volume
of the small gauge transformations,
the mismatch coming from the
ghosts for  ghosts phenomenon is the factor $(\det{}'\Delta_0)^{-1}$.
The volume of the large gauge transformations is cancelled by
restricting the integral over $\mathrm{Harm}^2(M_4,\Rh)$
to integral over the Jacobian  $\mathbb{T}=\mathrm{Harm}^2(M_4,\Rh)/\mathrm{Harm}^2_{\Zh}(M_4)$.
So after the the gauge fixing one obtains the following
expression for the norm:
\begin{equation}
\langle\Psi_{\beta},\Psi_{\beta'}\rangle=
\frac{\det{}'\Delta_1}{(\det{}'\Delta_0)^2}\int_{\mathbb{T}\times \mathbb{T}}
\Ds_g c^h \Ds_g b^h\,
\overline{\Psi_{\beta}(b^h,c^h)}\Psi_{\beta'}(b^h,c^h).
\label{norm}
\end{equation}

Now we substitute \eqref{psiPhys} into \eqref{norm}.
The field $c^h$ appears linearly in the exponential,
the integral over $c^h$ yields two Kronecker symbols $\delta_{\beta\beta'}$
and $\delta_{\omega,\omega'}$ where $\omega$ and $\omega'$
are summation variables in the definition of the $\Theta$-function:
\begin{equation*}
\langle\Psi_{\beta}^{\text{phys}},\Psi_{\beta'}^{\text{phys}}\rangle=
\delta_{\beta,\beta'}\,|\mathscr{N}(\tau)|^2\,\frac{\det{}'\Delta_1}{(\det{}'\Delta_0)^2}
\sum_{\omega\in \Lambda}
\int_{\mathbb{T}}\Ds_g b^h\,
e^{-2\pi N\tau_2\int_{M_4}
(b^h+\omega+\frac{1}{N}\beta)*(b^h+\omega+\frac{1}{N}\beta)
}.
\end{equation*}
The sum over $\omega$ combines with the integral over $\mathbb{T}$
to the integral over $\mathrm{Harm}^2(M_4,\Rh)$. This Gaussian
integral is easy to calculate, and one finally finds that
 the normalization constant $\mathscr{N}(\tau)$ is
\begin{equation}
\mathscr{N}(\tau)=(2N\tau_2)^{\mathrm{b}_2/4}\,\det{}'\Delta_0
(\det{}'\Delta_1)^{-1/2}.
\label{N(tau)}
\end{equation}
Notice that $\mathscr{N}(-1/\tau)=|\tau|^{-\mathrm{b}_2/2}\mathscr{N}(\tau)$.

\subsection{Representation of the duality group}

The $SL(2,\Zh)$ group is realized as follows:
\begin{subequations}
\begin{equation}
T:\quad \Psi^{\text{phys}}_{\beta}(c+b+\tfrac12
w_2,b;\tau+1)=e^{\frac{i\pi}{N}(\beta,\beta)
-i\pi(w_2,\beta)-\frac{i\pi N}{2}(w_2,b)}
\,\Psi^{\text{phys}}_{\beta}(c,b;\tau);
\end{equation}
\begin{equation}
S:\quad \Psi^{\text{phys}}_{\beta}(b,-c;-1/\tau)=
(-i\tau)^{\sigma/4}(i\bar{\tau})^{-\sigma/4}
\,
N^{-\mathrm{b}_2/2}\,\sum_{\beta'\in\Lambda/\Lambda_N}e^{-\frac{2\pi i}{N}(\beta,\beta')}\,\Psi^{\text{phys}}_{\beta'}(c,b;\tau).
\end{equation}
\end{subequations}
Here $\sigma=\mathrm{b}_2^+-\mathrm{b}_2^-$ is the Hirzebruch
signature, $N^{\mathrm{b}_2}$ is the order of the finite group  $\Lambda/\Lambda_N$. One sees that the physical wave functions
are modular forms of weight $(\sigma,-\sigma)$. Here $w_2$ is a
characteristic vector, such that
\begin{equation*}
(\omega,\omega)=(\omega,w_2)\mod 2.
\end{equation*}
For a spin manifold it equals zero.

\subsection{Dual conformal field theory}

Finally, one can interpret the $\Theta$-function in \eqref{psiPhys} as
a path integral over the $U(1)$ gauge field $A$ with
$F=dA$. We normalize $A$ such that $F$ has integral periods.
The dual action in the topological sector can be written as
\begin{subequations}
\begin{multline}
e^{iI_{\text{dual}}(F,\tau;\,c,b)}=
e^{i\pi N\int_{M_4}c\wedge b
-2\pi i N\int_{M_4}c\wedge(F+b+\frac{1}{N}\beta)
}
\\
\times
\exp\biggl[i\pi N \tau\int_{M_4}
(F+b+\tfrac{1}{N}\beta)_+^2
+i\pi N \bar{\tau}\int_{M_4}
(F+b+\tfrac{1}{N}\beta)_-^2
\biggr]
\label{dual}
\end{multline}
Here $\beta\in\mathrm{Harm}_{\Zh}^2(M_4)$ is a harmonic representative.
One sees that the $U(1)$ field $F$ is obtained
from the $U(1)$ part of the $U(N)$ gauge theory.
$\tau=\frac{\theta}{\pi}+\frac{i}{g_B}$ is the gauge theory coupling constant

An equivalent form, which makes the $SL(2,\Zh)$ properties manifest is:
\begin{multline}
e^{iI_{\text{dual}}(F,\tau;\,c,b)}=
e^{i\pi N \tau\int_{M_4}
(F+\frac{1}{N}\beta)_+^2
-2\pi i N\int_{M_4}(c-\tau b)_+\wedge(F+\frac{1}{N}\beta)_+
-i\pi N\int_{M_4}(c-\tau b)_+\wedge b_+
}
\\
\times
e^{i\pi N \bar{\tau}\int_{M_4}
(F+\frac{1}{N}\beta)_-^2
-2\pi i N\int_{M_4}(c-\bar{\tau} b)_-\wedge(F+\frac{1}{N}\beta)_-
-i\pi N\int_{M_4}(c-\bar{\tau}b)_-\wedge b_-
}.
\label{dual2}
\end{multline}
\end{subequations}
Note that $F_+$ transforms as a modular form of weight $1$
and hence must couple to $(c-\tau b)$ which transforms with weight $-1$.

It is very interesting that the normalization \eqref{N(tau)} of the bulk-theory wavefunction
is precisely the one-loop determinant of the gauge boson. Thus we confirm the AdS/CFT correspondence
at the full quantum level for this free field sector. It would be interesting to give a physical
interpretation to the wavefunction overlaps \eqref{inner} from the bulk supergravity viewpoint.

As we have stressed, we are assuming $M_4$ is spin in this paper.
If it is not spin then \eqref{dual} can be modified to make the theory
$SL(2,\Zh)$ invariant. One must shift the quantization law of $F$ by $w_2(M_4)$
and add a phase factor $\exp[i \pi \int F w_2(M_4)]$. Then the path integral
will be $SL(2,\Zh)$ invariant. \footnote{G.M. Thanks J. Evslin and
E. Witten for a useful conversation on this point.}
This indicates that the $B,C$ theory can
also be suitably modified to restore $SL(2,\Zh)$ invariance in the non-spin case.
We suspect that it is related to further subtleties in the IIB phase,
but we leave this for the future.

\section{'t Hooft and Wilson Lines}
\dopage{\finkfile}
\setcounter{equation}{0}
In this section we analyze how
the conformal blocks of the dual $U(1)$ gauge theory
changes in the presence of Wilson surfaces for $B$ and $C$ fields
 \cite{Witten:1998wy,Aharony:1998qu,Aharony:1999ti,Gross:1998gk}.
These are usually denoted
\begin{equation}
\exp[2\pi i \int_{\Sigma_B} B_2 - 2\pi i \int_{\Sigma_C} C_2]
\end{equation}
where $\Sigma_B$ and $\Sigma_C$ are $2$-manifolds in $X_5$.
We will denote this factor by
\begin{equation}
\mathrm{Hol}_{ B}(\Sigma_B;\gamma_B)\mathrm{Hol}^*_{ C }(\Sigma_C;\gamma_C),
\label{Holterm}
\end{equation}
where the  boundary of $\Sigma_B,\Sigma_C $ is a $1$-manifold $\gamma_B,\gamma_C$ in $M_4$.
The surfaces
$\Sigma_B, \Sigma_C$ need not be connected but for simplicity of notation we will
assume below that they, and $\gamma_B, \gamma_C$ are connected.
For a closed surface $\Sigma_B$ the holonomy is
a complex number, and the following variational formula holds:
\begin{equation}
\mathrm{Hol}_{B+b_2}(\Sigma_B)=\mathrm{Hol}_{B}(\Sigma_B)\,
e^{2\pi i\int_{\Sigma_B}b_2}
\label{varHol}
\end{equation}
where $b_2$ is a globally well defined $2$-form.
For   surfaces with   boundary the holonomy
is a section of a line bundle over
$\Omega^2(M_4)\times Z_1(M_4)$. This line bundle
is defined by the usual gluing law:
Let $(\tilde{B},\Sigma_B)$ and $(\tilde{B}',\Sigma_B')$ be two extensions
of the contour $\gamma_B\in Z_1(M_4)$
and  $B$-field $B\in\Omega^2(M_4)$
to $X_5$ then the holonomies are related by
\begin{equation}
\frac{\mathrm{Hol}_{\tilde{B}}(\Sigma_B;\gamma_B)}{\mathrm{Hol}_{\tilde{B}'}(\Sigma_B';\gamma_B)}
=\mathrm{Hol}_{\tilde{B}-\tilde{B}'}(\Sigma_B\bar{\Sigma}_B').
\end{equation}
The phase on the right hand side is the holonomy of the $B$-field around the closed
surface $\Sigma_B\bar{\Sigma}_B'$ which is obtained
by gluing $\Sigma_B$ and $\bar{\Sigma}_B'$ along the boundary.

Inclusion of the holonomies \eqref{Holterm} modifies
the line bundle in which wave functions lives.
The new line bundle is over the space $\Omega^2(M_4)\times\Omega^2(M_4)\times
Z_1(M_4)\times Z_1(M_4)$.
A connection on this line bundle is defined as follows:
we choose a path $p(t)=(C(t),B(t))$ and extensions
$\Sigma_C(t)=(\gamma_C(t),t)$, $\Sigma_B(t)=(\gamma_B(t),t)$ of the
loops $\gamma_C$ and $\gamma_B$, then define
\begin{equation}
U(p(t),\Sigma_B(t),\Sigma_C(t))=
\Phi(p(t);M_4\times[0,1])
\times\mathrm{Hol}_{B(t)}(\Sigma_B;\pd\Sigma_B)
\mathrm{Hol}_{C(t)}^*(\Sigma_C;\pd\Sigma_C).
\label{connection1}
\end{equation}
It is straightforward to compute the curvature
of \eqref{connection1}, the components along
 $\Omega^2(M_4)\times\Omega^2(M_4)$  are given by \eqref{Omegacurv}
as before, while the components along $Z_1(M) \times Z_1(M)$ are given by
$\int_{\gamma_B} H - \int_{\gamma_C} F$ (where $H,F$ are the fieldstrengths
of a family of 2-form connections over $Z_1(M) \times Z_1(M)$).

It is easy to see that for the straightline path \eqref{path}
the composition of the parallel transports
is given by Eq.~\eqref{Umult}.
We can still choose the cocycle \eqref{cocycle} to define
the group lift. To define the action of the gauge group
on the wave functions, we must first trivialize the
line bundle. To this end we first choose the reference point
$(C_{\bullet},B_{\bullet})$. Then any field in this
cohomology class can be written as $(C_{\bullet}+\bar{c},\,B_{\bullet}+\bar{b})$.
We also have to choose  base contours $(\gamma_C^{\bullet},
\gamma_B^{\bullet})$ which represent some homology class
in  $H_1(M_4,\Zh)\times H_1(M_4,\Zh)$. Then
an arbitrary element $(\gamma_C,\gamma_B)$ from the same homology class
is related to base curve by adding a $2$-chain $(D_c,D_b)$
where $(\pd D_c,\pd D_b)=(\gamma_C-\gamma_C^{\bullet},\gamma_B-\gamma_B^{\bullet})$.
Now we can proceed as in section~3.2, we choose a standard
nowhere vanishing section by the parallel transport.
Then the wave function $\psi$ is
defined as the ratio of  a section $\Psi$ and the standard section.
This   leads to the following modification of
the Gauss law:
\begin{equation}
\psi(\bar{c},\bar{b})=
e^{*}_{\omega_c,\omega_b}(\bar{c},\bar{b})
e^{-2\pi i\int_{D_c}\omega_c+2\pi i\int_{D_b}\omega_b}
\psi(\bar{c}+\omega_c,\bar{b}+\omega_b).
\label{newGS}
\end{equation}

As a cross check let us compare this Gauss law with
the classical one. Near the boundary the surfaces $\Sigma_B$
and $\Sigma_C$ look like direct products $\Rh_+\times \gamma_B$
and $\Rh_+\times\gamma_C$. It is easy to see that
the Gauss law \eqref{GaussLAW} is modified to:
\begin{equation}
\mathcal{G}_c=-d(\Pi_c+\pi N\bar{b})+2\pi\delta(\gamma_C)=0
\quad\text{and}\quad
\mathcal{G}_c=d(-\Pi_b+\pi N\bar{c})-2\pi\delta(\gamma_B)=0.
\end{equation}
Similarly to \eqref{ttact} this yields
\begin{equation*}
\psi(\bar{c},\bar{b})=
e^{i\pi N\int_{M_4}
d\alpha_c\wedge \bar{b}-
d\alpha_b\wedge \bar{c}
}\,e^{-2\pi i\int_{\gamma_C}\alpha_c+2\pi i\int_{\gamma_B}\alpha_b}\,
\psi(\bar{c}+d\alpha_c,\,\bar{b}+d\alpha_b)
\end{equation*}
which agrees with \eqref{newGS} for $\omega_c=d\alpha_c$ and
$\omega_b=d\alpha_b$.
Clearly the Hamiltonian near the boundary does not change in the presence
of Wilson surfaces. Therefore we can take the solution \eqref{Psi0}
and average it over the modified large gauge transformations
\eqref{newGS}.
Using the techniques presented in section~4 and appendix~B
one finds that the action for the  dual gauge theory is
\begin{multline}
e^{iI_{\text{dual}}(F,\tau;\,c,b;D_c,D_b)}=
e^{2\pi i \int_{D_b}(F+\frac{1}{N}\beta)}
\,
e^{-i\pi N\int_{M_4}c\wedge b
-2\pi i N\int_{M_4}c\wedge(F-\frac{1}{N}\delta(D_c)+\frac{1}{N}\beta)
}
\\
\times
\exp\biggl[i\pi N \tau\int_{M_4}
(F+b-\tfrac{1}{N}\delta(D_c)+\tfrac{1}{N}\beta)_+^2
+i\pi N \bar{\tau}\int_{M_4}
(F+b-\tfrac{1}{N}\delta(D_c)+\tfrac{1}{N}\beta)_-^2
\biggr].
\label{dualJ}
\end{multline}
Here the action includes the Wilson line for $A$, written as a surface
integral over $D_b$, as expected. The presence of $\delta$-functions indicates
the need for some regularization of the self-energy of the
't~Hooft operators.
Making precise sense of this factor lies beyond the
scope of this paper. We can, however, confirm the dependence on the
choice of trivialization observed in \cite{Witten:1998wy} as follows.

As   explained above, a trivialization
is related to a choice of $D_c$ and $D_b$.
Let $D_c'$ be a different $2$-chain such that $\pd D_c'=\gamma_C-\gamma_C^{\bullet}$.
Consider a $2$-cycle $E=D_c'-D_c$. We want to make a change of
variable in the path integral of the dual gauge theory \eqref{dualJ}
such that
$F\mapsto F-\tfrac{1}{N}\delta(E)$.
We, certainly, can do this if $E$ is a boundary of a $3$-manifold $Y$,
then we just shift $A$ by $-\tfrac{1}{N}\theta(Y)$
where $\theta(Y)$ is the characteristic function of $Y$.
This change of integration variable not only changes $D_c$ to $D_c'$
in \eqref{dualJ} but also yields an extra factor:
\begin{equation}
\int \Ds A\,e^{iI_{\text{dual}}(F,\tau;c,b;D_c,D_b)}
=
e^{-\frac{2\pi i}{N}\int_{D_b}\delta(E)}
\,\int \Ds A\,e^{iI_{\text{dual}}(F,\tau;c,b;D_c',D_b)}.
\end{equation}
The integral in the first term on the right hand side is an integer which
equals the intersection number
$\#(D_c\cdot E)$ of
the $2$-chain $D_b$ with the $2$-cycle $E$.
Hence, under a change of trivialization the expectation value
of a product of both 't~Hooft and Wilson lines is multiplied
by an  $N^{\text{th}}$ root of unity in accord with \cite{Witten:1998wy}.

\section*{Acknowledgements}

We would like to thank E. Witten for discussions.
This work was supported  in part by DOE grant DE-FG02-96ER40949.

\appendix
\section*{Appendix}
\addcontentsline{toc}{section}{Appendix}
\renewcommand {\theequation}{\thesection.\arabic{equation}}

\section{Formulation in terms of Cheeger-Simons characters }
\label{app:CS}
\setcounter{equation}{0}

In this appendix we indicate briefly how the above paper could be formulated in
somewhat more abstract terms using Cheeger-Simons cohomology. Some discussion of
this can also be found in \cite{Witten:1998wy} section 3.5. The advantage of
this formalism is that it also covers the case when the cohomology $H^*(M_4,\Zh)$
has torsion.

The gauge invariant information in the supergravity potentials $B_2, C_2$ is properly regarded as
a Cheeger-Simons character $\check B, \check C \in \check H^3(M_4)$. For recent reviews of
Cheeger-Simons characters in this context see
\cite{HS,Freed,Diaconescu:2003bm}.
We follow the notation of \cite{Diaconescu:2003bm}.
If $X_5$ is oriented and compact there
 is a canonical multiplication   and integration and the expression
\begin{equation}
\exp[ 2\pi i N \int_{X_5} \check B \cdot  \check C ]
\label{cheegi}
\end{equation}
is a well-defined phase. When $X_5$ is
oriented with boundary $M_4$ \eqref{cheegi} must be regarded as a section of a
line bundle $\mathcal{L}_N$
over the space
\begin{equation}
 \check H^3(M_4)\times \check H^3(M_4)
\label{base}
\end{equation}
Moreover, $\mathcal{L}_N$ comes together with a canonical connection. (See for example
the discussion in section 3.3 above.)  There is an Hermitian metric on $\mathcal{L}_N$
so that the connection is unitary. The connection has curvature
\begin{equation}
\Omega = 2\pi N \int_{M_4} \delta B_2 \wedge \delta C_2
\label{curv}
\end{equation}
and is $N$ times the canonical symplectic form on \eqref{base}.
However, since we consider the theory with standard kinetic terms the
 wavefunctions in the quantum theory
are   in the Hilbert space
\begin{equation}
L^2\bigl( \check H^3(M_4)\times \check H^3(M_4); \mathcal{L}_N)
\label{hilbspace}
\end{equation}
and the Hamiltonian is the canonical Laplacian  where we use the
$SL(2,\Zh)$-covariant, translation invariant, metric $(B.3)$ below.

Note that the main difference between \cite{Witten:1998wy} and the present discussion is that
in \cite{Witten:1998wy}, \eqref{base} was considered to be a phase space. In the present discussion
it is the configuration space, and the phase space is the cotangent bundle of \eqref{base}.

Now, the space $\check H^3(M_4)\times \check H^3(M_4)$ is a disjoint union of
spaces modelled on $\Omega^2/\Omega^2_{\Zh} \times \Omega^2/\Omega^2_{\Zh}$, with the
connected components labelled by $H^3(M,\Zh)\oplus H^3(M,\Zh)$. At this point
we encounter an interesting subtlety. The space \eqref{base} actually labels the
isomorphism classes of a field with automorphisms, as described in
\cite{HS,Diaconescu:2003bm,mooreS04}.
The gauge fields are objects in a   groupoid, and the automorphism group of an object is
$H^1(M_4,U(1))\times H^1(M_4,U(1))$. In order for the line bundle $\mathcal{L}_N$ to be well-defined
on \eqref{base} we require the tadpole condition on the characteristic classes,
determined by the requirement that the automorphism group act trivially on $\mathcal{L}_N$.
The action of this automorphism group is given as follows. Suppose $(\check \chi_c, \check \chi_b)$
are flat characters in $\check  H^2(M_4) \times \check  H^2(M_4)$, and let $\check  t$ be the canonical
character on $\check  H^1(S^1)$. Then the ``lifting phase'' of section 3.2 is properly defined by
\begin{equation}
\varphi(\check  C, \check  B; \check \chi_c, \check \chi_b)
= \Phi_B( \check  B + \check  t \cdot \check  \chi_b, \check  C + \check  t\cdot \check  \chi_c; M_4\times S^1)
\end{equation}
When restricted to flat characters this is a homomorphism, and if $\check  B, \check  C =0$ then it is Poincar\'e dual to
torsion background charges $\mu_c, \mu_c \in H^3_T(M_4;\Zh)$, analogous to the class $\mu$ discussed in \cite{HS}
or the class $\Theta(0)$ discussed in \cite{Diaconescu:2003bm}.
The Gauss law on the characteristic class is
\begin{equation}
a(\check B) =\mu_b\qquad\qquad  a(\check C) = \mu_c.
\label{tadpole}
\end{equation}
where $a(\check  B)$ denotes the characteristic class of the Cheeger-Simons character.

The condition \eqref{tadpole} is  analogous to the tadpole
 condition for the $M$-theory $3$-form, given by  equation $(7.7)$ of
\cite{Diaconescu:2003bm}, and it arises in the same way.
To explain this, let   us note parenthetically that it   is possible to give an analog of the
``$E_8$ model for the $C$-field''
for  2-form potentials whose isomorphism class is an element of $\check H^3(M)$.
\footnote{This formulation suggests a speculation. This model could be
applied to the 2-form gauge field on the $M5$ brane, thus reformulating the 5-brane
theory as a six-dimensional nonlinear sigma model with target space $E_8$. It is then
natural to ask if the map to $E_8$ ``becomes dynamical'' for coincident 5-branes,
in a way analogous to the way the topological $E_8$ gauge field of $M$-theory
becomes a dynamical gauge field in heterotic $M$-theory.   }
Let $G$ be a compact Lie group whose homotopy type is that of $K(\Zh,3)$ up to
the $n$-skeleton. (For example, $G=E_8$ for $n<15$.) On a manifold of
dimension $n$ we define an object in the groupoid to be a pair $(g,b)$ where
$g:M \to G$ is a smooth map and $b\in \Omega^2(M)$ is a globally defined $2$-form.
The isomorphism class of $(g,b)$ is the differential character defined by the
holonomies
\begin{equation}
\check \chi_{g,b}(\Sigma) = e^{2\pi i (WZ(g,\Sigma) + \int_{\Sigma} b )}
\end{equation}
Here $WZ(g,\Sigma)$ is the Wess-Zumino term, thus $WZ(g,\Sigma) = \int_{B} \Tr(g^{-1} dg)^3 $
where $\partial B = \Sigma$ and $\Tr(g^{-1} dg)^3$ is the pullback of a representative
of a generator of $H^3(G,\Zh)$. The field strength is
$\omega(\check \chi_{g,b} ) = \Tr(g^{-1} dg)^3 + db$ and the characteristic class is
the homotopy class of $g: M \to G$. We will not give the full description of the morphisms
here. It  suffices to note that the automorphism group of an object is $H^1(M,U(1))$.

The connection on ${\cal L}_N$  is $SL(2,\Zh)$ invariant  and thus $SL(2,\Zh)$ acts on
the Hilbert space, as described in \cite{Witten:1998wy}.
The translation group of
\eqref{base} on itself is not an invariance of the connection,
thus leading to the action of the
magnetic translation group, described by the Page charges.
This is the Heisenberg group
described in equation \eqref{Hgroup} above.  The space
$\Omega^2/\Omega^2_{\Zh} \times \Omega^2/\Omega^2_{\Zh}$ contains the
torus $H^2(M,\Rh/\Zh) \times H^2(M,\Rh/\Zh)$,
and the theta functions in this paper define the appropriate
sections of $\mathcal{L}_N$ over this torus.
The relevant complex structure and polarization are described in appendix B below.

\section{Gaussian sums on $\mathrm{Harm}^2(M_4,\Zh)\times
\mathrm{Harm}^2(M_4,\Zh)$}
\label{app:GS}
\setcounter{equation}{0}
\subsection{Symplectic structure, complex structure, and metric}
Consider the lattice $V_{\Zh}=\mathrm{Harm}^2_{\Zh}(M_4)
\times \mathrm{Harm}^2_{\Zh}(M_4)$ of   rank $2\mathrm{b}_2(M_4)$.
This lattice has integral valued symplectic form $\Omega$
\begin{equation}
\Omega\bigl((\omega_c,\omega_b),(\omega_c',\omega_b')\bigr)
=\int_{M_4}
\begin{pmatrix}
\omega_c & \omega_b
\end{pmatrix}
\begin{pmatrix}
0 & -1
\\
1 & 0
\end{pmatrix}
\begin{pmatrix}
\omega_c' \\
\omega_b'
\end{pmatrix}.
\end{equation}
Given a metric on $M_4$ and a complex number $\tau$ with $\tau_2>0$
one can define a complex structure $J$ on $V_{\Rh}$:
\begin{equation}
J\begin{pmatrix}
\phi_c
\\
\phi_b
\end{pmatrix}
=\begin{pmatrix}
0 & -1
\\
1 & 0
\end{pmatrix}
\Mc(\tau)
\begin{pmatrix}
*\phi_c
\\
*\phi_b
\end{pmatrix}
\label{J}
\end{equation}
where $\Mc(\tau)$ is defined in \eqref{M(tau)}.
It is easy to see that $J^2=-1$. This complex structure is compatible
with the symplectic one
\begin{equation*}
\Omega\bigl(J\cdot(\phi_c,\phi_b),\,
J\cdot(\phi_c',\phi_b')\bigr)=\Omega\bigl((\phi_c,\phi_b),\,
(\phi_c',\phi_b')\bigr).
\end{equation*}
In this case we can define a quadratic form
\begin{equation}
g\bigl((\phi_c,\phi_b),\,
(\phi_c',\phi_b')\bigr)
=\Omega\bigl(J\cdot(\phi_c,\phi_b),\,(\phi_c',\phi_b')\bigr)
=\int_{M_4}(*\phi_c,*\phi_b)\Mc(\tau)
\begin{pmatrix}
\phi_c'
\\
\phi_b'
\end{pmatrix}
\end{equation}

We can choose a symplectic basis $\alpha^I,\beta_I$ for $V_{\Zh}$ to be
\begin{equation}
\alpha^I=(0,\,\omega^I)\quad\text{and}\quad
\beta_I=(\hat{\omega}_I,\,0)
\label{sbasis}
\end{equation}
where $\omega^I$ and $\hat{\omega}_I$ are dual bases in $\mathrm{Harm}^2_{\Zh}(M_4)$.
If $\boldsymbol{\tau}^{IJ}$ is the period matrix in the basis
$\omega^I$ then $\hat{\omega}_I\boldsymbol{\tau}^{IJ}=\omega^J$.
One can verify that basis \eqref{sbasis} is indeed a symplectic basis
\begin{equation*}
\Omega(\alpha^I,\alpha^J)=0=\Omega(\beta_I,\beta_J)
\quad\text{and}\quad
\Omega(\alpha^I,\beta_J)=\delta^I{}_J.
\end{equation*}

The complex structure \eqref{J} acts on the basis \eqref{sbasis} as
follows
\begin{equation}
J\cdot\begin{pmatrix}
\alpha^I
\\
\beta_I
\end{pmatrix}
=\frac{1}{\tau_2}
\begin{pmatrix}
-\tau_1(h\boldsymbol{\tau}^{-1})^I{}_J & -|\tau|^2 h^{IJ}
\\
(h^{-1})_{IJ} &
\tau_1(h^{-1}\boldsymbol{\tau})_I{}^J
\end{pmatrix}
\begin{pmatrix}
\alpha^J
\\
\beta_J
\end{pmatrix}
\label{mJ}
\end{equation}
where
\begin{equation*}
h^{IJ}=\int_{M_4}\omega^I *\omega^J\quad\text{and}\quad
\boldsymbol{\tau}^{IJ}
=\int_{M_4}\omega^I \wedge\omega^J.
\end{equation*}

Now choose the basis $\zeta^I$ of type $(1,0)$. By definition
$\zeta^I$ is a basis of solutions of the equation
$J\cdot\zeta^I=i\zeta^I$.
One can express the complex structure $J$ in terms
of the components of the complex period matrix $\mathbf{T}$.
To this end we choose a basis $\zeta^I$ of the form
\begin{equation}
\zeta^I=\alpha^I+\mathbf{T}^{IJ}\beta_J.
\end{equation}
{}From $g(\zeta^I,\zeta^J)=g(\zeta^J,\zeta^I)$ we learn that
$\mathbf{T}^{IJ}$ is symmetric, and $g$ is of type $(1,1)$.
Note that $g(\zeta^I,\bar{\zeta}^J)=2\Im\mathbf{T}^{IJ}$.
{}From Eq.~\eqref{mJ} one finds
\begin{equation}
\mathbf{T}^{IJ}=\tau_1\boldsymbol{\tau}^{IJ}+i\tau_2 h^{IJ}.
\label{T}
\end{equation}

We write $\mathbf{T}=X+iY$ and
\begin{equation}
J\cdot\begin{pmatrix}
\alpha^I
\\
\beta_I
\end{pmatrix}
=
\begin{pmatrix}
-(XY^{-1})^I{}_J & -(Y+X
Y^{-1}X)^{IJ}
\\
(Y^{-1})_{IJ} &
(Y^{-1}X)_I{}^J
\end{pmatrix}
\begin{pmatrix}
\alpha^J
\\
\beta_J
\end{pmatrix}
\label{mTJ}
\end{equation}
Let $\nu=n_I\alpha^I+m_I\beta^I$
and $\tilde{\nu}=\tilde{n}_I\alpha^I+\tilde{m}^I\beta_I$,
then the metric in the $\alpha,\beta$ basis is:
\begin{equation}
g(\nu,\tilde{\nu})=
\begin{pmatrix}
n_I
&
m^I
\end{pmatrix}
\begin{pmatrix}
(Y+X
Y^{-1}X)^{IJ} &
-(XY^{-1})^I{}_J
\\
-(Y^{-1}X)_I{}^J &
(Y^{-1})_{IJ}
\end{pmatrix}
\begin{pmatrix}
\tilde{n}_J
\\
\tilde{m}^J
\end{pmatrix}.
\label{gnn}
\end{equation}

\subsection{Splitting the instanton sum}
Define level $k$ theta functions to be
\begin{equation*}
\Theta_{\beta,k}(\xi,\mathbf{T})
=\sum_{s_I\in\Zh}\exp\Bigl[2\pi i k \bigl(s_I+\tfrac{1}{2k}\beta_I\bigr)
\mathbf{T}^{IJ}
\bigl(s_J+
\tfrac{1}{2k}\beta_J\bigr)
\Bigr]e^{2\pi i\xi^I(2k s_I+\beta_I)}.
\end{equation*}
Here $\mathbf{T}^{IJ}$ is the complex period matrix \eqref{T}.
We can rewrite this expression in a geometrical form. Define,
\begin{equation*}
s=s_I\omega^I,\qquad \beta=\beta_I\omega^I,\qquad
\xi=\xi^I\,\hat{\omega}_I.
\end{equation*}
Then the $\Theta$-function is
\begin{equation}
\Theta_{\beta,k}(\xi;\tau,*)
=\sum_{s\in\mathrm{Harm}^{2}_{\Zh}(M_4)}
e^{2\pi i k\tau\int_{M_4}(s+\frac{1}{2k}\beta)_+^2
+2\pi i k\bar{\tau}\int_{M_4}
(s+\frac{1}{2k}\beta)_-^2
+2\pi i\int_{M_4}\xi\wedge(2k s+\beta)}
.
\label{Thetak}
\end{equation}

We want to express
\begin{equation}
S:=\sum_{\nu\in V_{\Zh}}\varphi(\nu)e^{-\frac12\pi Ng(\nu,\nu)
+\Omega(\nu,\tilde{\ell})}
\label{sum1}
\end{equation}
where $g$ is defined in \eqref{gnn}
in terms of theta functions for the complex torus $V_{\Rh}/V_{\Zh}$.
Here $\nu=n_I\alpha^I+m^I\beta_I$, $\tilde{\ell}=-\ell_I^2\alpha^I+\ell_1^I\beta_I$,
therefore $\Omega(\nu,\tilde{\ell})=n_I\ell_1^I+m^I\ell^2_I$.
$\varphi(\nu)$ is a quadratic refinement of $\Omega$, i.e.
\begin{equation*}
\varphi(\nu_1+\nu_2)=\varphi(\nu_1)\varphi(\nu_2)e^{i\pi N\Omega(\nu_1,\nu_2)}.
\end{equation*}
For our purpose it
sufficient to consider $\varphi(\nu)$ of the form $\varphi(\nu)=e^{+i\pi N n_Im^I}$.
Geometrically the sum \eqref{sum1} is
\begin{equation}
S=\sum_{\omega_c,\omega_b\in\mathrm{Harm}^2_{\Zh}(M_4)}
e^{-\frac{\pi N}{2\tau_2}\int_{M_4}(\omega_c-\bar{\tau}\omega_b)
*(\omega_c-\tau\omega_b)+i\pi N\int_{M_4}\omega_c\wedge\omega_b}\,
e^{+\int_{M_4}\omega_b\wedge\ell_1+\omega_c\wedge\ell^2}.
\label{S_1forms}
\end{equation}

Now we do Poisson resummation over $m^I$:
\begin{equation}
S(\ell)=\Bigl(\frac{2\tau_2}{N}\Bigr)^{\mathrm{b}_2/2}
e^{\frac{1}{2\pi N}\ell_I^2Y^{IJ}\ell^2_J}\,
\sum_{n_I,w_I}
e^{
i\pi N(p_L)_I\mathbf{T}^{IJ}(p_L)_J
-i\pi N(p_R)_I\mathbf{\bar{T}}^{IJ}(p_R)_J
+(p_L)_I\psi^I+(p_R)_I\bar{\psi}^I}
.
\label{S1'}
\end{equation}
Here
\begin{equation}
(p_L)_I=\frac12\, n_I+\frac{1}{N}\,\Bigl(w_I+\frac{N}{2} n_I\Bigr);
\quad
(p_R)_I=\frac12\, n_I-\frac{1}{N}\,\Bigl(w_I+\frac{N}{2} n_I\Bigr);
\quad
\psi^I=\ell_1^I+T^{IJ}\ell^2_J.
\label{psiell}
\end{equation}
Now we write $w_I=\beta_I-Ns_I$ where  $\beta_I\in\{0,1,\dots,N-1\}$
and $s_I\in\Zh$. In this case
\begin{equation}
(p_L)_I=n_I-s_I+\frac{1}{N}\beta_I
\quad\text{and}\quad
(p_R)_I=s_I-\frac{1}{N}\beta_I.
\end{equation}
Finally the sum \eqref{S1'} takes the form
\begin{equation}
S(\ell)=\Bigl(\frac{2\tau_2}{N}\Bigr)^{\mathrm{b}_2/2}
e^{\frac{1}{2\pi N}\ell_I^2Y^{IJ}\ell^2_J}\,
\sum_{\beta\in(\Zh/N\Zh)^{\mathrm{b}_2}}
\Theta_{\beta,N/2}\Bigl(\frac{1}{2\pi i N}\psi^I,\,\mathbf{T}^{IJ}\Bigr)
\Theta_{-\beta,N/2}\Bigl(\frac{1}{2\pi i N}\bar{\psi}^I,\,-\mathbf{\bar{T}}^{IJ}\Bigr).
\label{splitting}
\end{equation}

\subsection{Summary}
The final result is that the Gaussian sum
\begin{multline}
S=\sum_{\omega_c,\omega_b\in\Lambda}
e^{-\frac{\pi N}{2\tau_2}\int_{M_4}(\omega_c-\bar{\tau}\omega_b)
*(\omega_c-\tau\omega_b)+i\pi N\int_{M_4}\omega_c\wedge\omega_b}
\\
\times
e^{\frac{1}{2i\tau_2}\int_{M_4}
[
\psi_+\wedge(\omega_c-\bar{\tau}\omega_b)_+-\psi_-\wedge(\omega_c-\tau\omega_b)_-
-\bar{\psi}_+\wedge(\omega_c-\tau\omega_b)_+
+\bar{\psi}_-\wedge(\omega_c-\bar{\tau}\omega_b)_-
]
}
\label{S_1forms}
\end{multline}
where $\Lambda=\mathrm{Harm}^2_{\Zh}(M_4)$
can be written as
\begin{equation}
S=\Bigl(\frac{2\tau_2}{N}\Bigr)^{\mathrm{b}_2/2}
e^{-\frac{1}{8\pi N\tau_2}\int_{M_4}(\psi-\bar{\psi})*(\psi-\bar{\psi})}
\sum_{\beta\in\Lambda/\Lambda_N}
\Theta_{\beta,N/2}\Bigl(\frac{1}{2\pi i N}\psi;\,\tau,*\Bigr)
\Theta_{-\beta,N/2}\Bigl(\frac{1}{2\pi i N}\bar{\psi};-\bar{\tau},*\Bigr).
\label{formTheta}
\end{equation}
where $\Lambda_N=\mathrm{Harm}^2_{N\Zh}(M_4)$.

For the sum \eqref{omegasum} we have
\begin{align*}
\psi_+&=-2\pi i N(c-\tau b)_+,\qquad \psi_-=-2\pi i N(c-\bar{\tau} b)_-;
\\
\bar{\psi}_+&=2\pi i (v_c-\bar{\tau} v_b)_+,\qquad\quad \bar{\psi}_-=2\pi i
(v_c-\tau v_b)_-.
\end{align*}
For these specific values of $\psi$ and $\bar{\psi}$
the sum \eqref{formTheta} can be rewritten in terms
of the Siegel-Narain $\Theta$-function defined by Eq.~\eqref{SNth}.
It has the following modular properties ($\gamma\in\frac{1}{2k}\Lambda$):
\begin{subequations}
\begin{equation}
T:\quad \Theta_{\Lambda+\gamma,\,k}(\tau+1,c+b+\tfrac12
w_2,b)=e^{2\pi ik(\gamma,\gamma)
-2\pi i k(w_2,\gamma)-i\pi k(w_2,b)}
\,\Theta_{\Lambda+\gamma,\,k}(\tau,c,b);
\end{equation}
\begin{equation}
S:\quad \Theta_{\Lambda+\gamma,\,k}(-1/\tau,b,-c)=
\frac{(-i\tau)^{\mathrm{b}_2^+/2}(i\bar{\tau})^{\mathrm{b}_2^-/2}}{
(2k)^{\mathrm{b}_2/2}}\,\sum_{\gamma'\in(\frac{1}{2k}\Lambda)/\Lambda}
e^{-4\pi i k(\gamma,\gamma')}\,\Theta_{\Lambda+\gamma',\,k}(\tau,c,b).
\end{equation}
\end{subequations}
Here $w_2$ is a
characteristic vector, such that
\begin{equation*}
(\omega,\omega)=(\omega,w_2)\mod 2.
\end{equation*}
For a spin manifold it equals zero.


\end{document}